\documentclass[11pt, letter]{article}
\usepackage{jheppub}
\usepackage[english]{babel}
\usepackage{blindtext}\blindmathtrue
\usepackage{avant} 
\usepackage[dvipsnames]{xcolor}
\usepackage{amsmath,epsf,amssymb,mathtools,latexsym,amsthm,setspace,array,pifont,hyperref,amsfonts,dsfont,cancel,braket,parskip}
\usepackage[none]{hyphenat} 
\usepackage{graphicx}
\usepackage{float}
\usepackage{tikz}
\usepackage{tikz-feynman, contour}
\tikzfeynmanset{compat=1.1.0}
\usetikzlibrary{shapes,arrows,positioning,automata,backgrounds,calc,er,patterns}
\makeatletter
\newlength\CoolS@sizex
\newlength\CoolS@sizey
\newcommand*\CoolS@inner{%
	\begin{tikzpicture}[baseline=0.04\CoolS@sizey]%
		\foreach \x in {0, 1, ..., 5} \foreach \y in {0, 1, ..., 10}
		\coordinate (c\x\y) at (\x *0.12*\CoolS@sizex, \y *0.107*\CoolS@sizey);
		\draw [line width=\Cool@stroke] (c28)--(c26)--(c44)--(c42)--(c20)--(c02)--(c04)--(c15);
		\draw [line width=\Cool@stroke] (c22)--(c24)--(c06)--(c08)--(c210)--(c48)--(c46)--(c35);
\end{tikzpicture}}
\usepackage{todonotes}
\graphicspath{{Pictures/}} 

\newcommand{\nn}{\nonumber}
\newcommand{\sd}{\mathrm{d}}
\newcommand{\pd}{\partial}

\newcommand{\R}{\mathbb{R}}

\newcommand{\bb}[1]{\mathbb{#1}}
\newcommand{\cl}[1]{\mathcal{#1}}

\renewcommand{\Re}{\operatorname{Re}}

\def\prd{\ref@{Phys.~Rev.~D}}        

\renewcommand{\[}{\begin{equation}\begin{aligned}}
		\renewcommand{\]}{\end{aligned}\end{equation}}

\usepackage{tcolorbox}
\definecolor{airforceblue}{rgb}{0.36, 0.54, 0.66}
\definecolor{azure}{rgb}{0.0, 0.5, 1.0}
\newtcolorbox{tdbox}{colback=airforceblue!40!white,colframe=azure!90!black} 
\newcommand{\td}[1]{
	\if\notesOn1
	\begin{tdbox}
		#1
	\end{tdbox}
	\fi
}
\hypersetup{
	colorlinks = true,
	linkcolor = Mahogany,
	anchorcolor = Mahogany,
	citecolor = Mahogany,
	filecolor = Mahogany,
	urlcolor = Mahogany
}

\def\notesOn{1}
\tikzset{
	graviton/.style={
		double,
		decoration={snake, aspect=0.75, mirror, segment length=1.5mm},
		decorate
	}
}

\tikzfeynmanset{
	bigblob/.style={
		shape=circle,
		draw=blue,
		fill=red}
}

\preprint{QMUL-PH-22-38, Imperial/TP/2022/MC/06}

\title{Mini-twistors and the Cotton Double Copy}
\author[1] {Mariana Carrillo Gonz\'{a}lez,}
\author[2]{William~T.~Emond,}
\author[3]{Nathan Moynihan,}
\author[5] {Justinas Rumbutis}
\author[4] {and Chris D. White}
\affiliation[1]{Theoretical Physics, Blackett Laboratory, Imperial College, London, SW7 2AZ, United Kingdom}
\affiliation[2]{CEICO, Institute of Physics of the Czech Academy of Sciences, Na Slovance 2, 182 21 Praha 8, Czech Republic}
\affiliation[3]{Higgs Centre for Theoretical Physics, School of Physics and Astronomy, The University of Edinburgh, EH9 3FD, Scotland, United Kingdom}
\affiliation[4]{Department of Physics and Astronomy, Queen Mary University of London, 327 Mile End Road, London, E1 4NS, United Kingdom}
\affiliation[5]{Department of Physics and Center for Theoretical Physics, National Taiwan University, Taipei
10617, Taiwan}
\emailAdd{m.carrillo-gonzalez@imperial.ac.uk, william.emond@fzu.cz, nathantmoynihan@gmail.com, justinasru@ntu.edu.tw, christopher.white@qmul.ac.uk}
\abstract{
	The double copy relates quantities in gauge, gravity and related theories. A well-known procedure for relating exact classical solutions is the Weyl double copy in four spacetime dimensions, and a three-dimensional analogue of this -- the Cotton double copy -- has recently been found for topologically massive gauge theory and gravity. In this paper, we use twistor methods to provide a derivation of the position-space Cotton double copy, where this is seen to arise from combining appropriate data in so-called minitwistor space. Our methods rely on a massive generalisation of the Penrose transform linking spacetime fields with cohomology classes in minitwistor space. We identify the relevant transform from the twistor literature, but also show that it naturally arises from considering scattering amplitudes in momentum space. We show that the Cotton double copy in position space is only valid for type N solutions, but that a simple twistor space double copy is possible for non-type N solutions, where we use anyons to illustrate our arguments.
}

\begin{document}
\maketitle

\section{Introduction}

In recent years, a correspondence known as the {\it double copy} has generated a great deal of interest. Inspired by previous work in string theory~\cite{Kawai:1985xq}, its original incarnation stipulates that scattering amplitudes in gauge theory can be straightforwardly turned into gravity amplitudes~\cite{Bern:2008qj,Bern:2010ue,Bern:2010yg}. To do so, one must substitute the appropriate coupling constants, as well as replace colour charge information with additional kinematic factors. This works for a wide variety of gauge and gravity theories, both with and without supersymmetry. Furthermore, one may also start with gauge amplitudes and go the other way, replacing kinematic with colour information. This is called the {\it zeroth copy}, and generates amplitudes in a scalar theory with two distinct types of colour charge, which has become known as {\it biadjoint scalar theory}. Whilst not a physical theory by itself, its dynamics is at least partially inherited by gauge and gravity theories. Furthermore, this ladder of theories includes a wide variety of examples (e.g. both with and without supersymmetry), and is itself part of a wider web of theories known to exhibit such correspondences: see e.g. refs.~\cite{Bern:2019prr,Borsten:2020bgv,Adamo:2022dcm,Bern:2022wqg} for recent reviews. In the past few years, it has become increasingly recognised that the double copy applies beyond fixed-order scattering amplitudes, in particular to all-order perturbative information~\cite{Oxburgh:2012zr,Vera:2012ds,Johansson:2013nsa,Saotome:2012vy}, exact classical solutions~\cite{Monteiro:2014cda,Luna:2015paa,Ridgway:2015fdl,Bahjat-Abbas:2017htu,Carrillo-Gonzalez:2017iyj,CarrilloGonzalez:2019gof,Bah:2019sda,Alkac:2021seh,Alkac:2022tvc,Luna:2018dpt,Sabharwal:2019ngs,Alawadhi:2020jrv,Godazgar:2020zbv,White:2020sfn,Chacon:2020fmr,Chacon:2021wbr,Chacon:2021hfe,Chacon:2021lox,Dempsey:2022sls,Easson:2022zoh,Chawla:2022ogv,Han:2022mze,Armstrong-Williams:2022apo,Han:2022ubu} (see also refs.~\cite{Didenko:2008va,Didenko:2009td} for related work in a different context, and~\cite{Didenko:2022qxq} for a recent overview of how this is related), perturbative classical solutions~\cite{Elor:2020nqe,Farnsworth:2021wvs,Anastasiou:2014qba,LopesCardoso:2018xes,Anastasiou:2018rdx,Luna:2020adi,Borsten:2020xbt,Borsten:2020zgj,Goldberger:2017frp,Goldberger:2017vcg,Goldberger:2017ogt,Goldberger:2019xef,Goldberger:2016iau,Prabhu:2020avf,Luna:2016hge,Luna:2017dtq,Cheung:2016prv,Cheung:2021zvb,Cheung:2022vnd,Cheung:2022mix}, and potential non-perturbative aspects~\cite{Monteiro:2011pc,Borsten:2021hua,Alawadhi:2019urr,Banerjee:2019saj,Huang:2019cja,Berman:2018hwd,Alfonsi:2020lub,Alawadhi:2021uie,White:2016jzc,DeSmet:2017rve,Bahjat-Abbas:2018vgo,Cheung:2022mix,Borsten:2022vtg}. 

The double copy offers new calculational tools for General Relativity and related theories, and indeed has already been used to generate new results needed for gravitational wave experiments~\cite{Bern:2019nnu,Bern:2019crd}. However, it also offers new conceptual insights not only about gravity, but also the very foundations of field theory itself. It is then important to find explanations of where the double copy comes from, particularly in those cases in which exact statements can be made. In four spacetime dimensions, a well-known exact classical double copy is the {\it Weyl double copy} of ref.~\cite{Luna:2018dpt}. Written using the spinorial formalism of field theory (see e.g. refs.~\cite{Penrose:1985bww,Penrose:1986ca,Stewart:1990uf} for reviews), it relates spacetime fields in biadjoint, gauge and gravity theories directly in position space\footnote{More specifically, it relates the Abelian versions of these theories. For explorations where both Abelian and non-Abelian solutions map to the same gravitational object see \cite{Bahjat-Abbas:2020cyb,Oxburgh:2012zr}.}. This is at odds with the original double copy for scattering amplitudes~\cite{Bern:2008qj,Bern:2010ue,Bern:2010yg}, which is naturally formulated in momentum space. Nevertheless, it is argued to be exact for certain gravity solutions, namely those vacuum solutions which are of type D in the well-known {\it Petrov classification}. Exact type N cases are also known~\cite{Godazgar:2020zbv}, as well as other Petrov types at linearised level only~\cite{White:2020sfn,Chacon:2021wbr}. 

In order to explain the above results, refs.~\cite{White:2020sfn,Chacon:2021wbr} found a derivation of the Weyl double copy using twistor theory~\cite{Penrose:1967wn,Penrose:1968me,Penrose:1972ia}, a decades-old set of mathematical ideas linking field theory, complex analysis and algebraic geometry (see e.g. refs.~\cite{Penrose:1986ca,Huggett:1986fs,Adamo:2017qyl} for pedagogical reviews). In a nutshell, twistor theory maps points in spacetime non-locally to an abstract {\it twistor space}, such that certain quantities in the former show up as mathematically convenient data in the latter. In particular, an integral formula known as the {\it Penrose transform}~\cite{Penrose:1969ae} relates certain ``functions" in twistor space to spacetime solutions of the massless free field equations, where a given spin of the spacetime field translates to a given homogeneity of the twistor function under rescalings of its argument. By multiplying together functions of different homogeneity in an appropriate manner, refs.~\cite{White:2020sfn,Chacon:2021wbr} showed that one could derive the Weyl double copy in position space. Furthermore, the twistor approach provided a geometric interpretation of certain aspects of the Weyl double copy that had previously been obscure, such as the {\it inverse zeroth copy} that takes one from biadjoint scalar to gauge theory. Applications of the twistor approach include generalising the Weyl double copy away from type D or N solutions (albeit at linearised level only), and also showing that physical properties such as multipoles can be straightforwardly mapped between different theories~\cite{Chacon:2021hfe}. 

Despite the above successes, the twistor double copy is not without its conceptual problems. Chief among these is the fact that the quantities entering the Penrose transform are not, strictly speaking, functions. Rather, they may be redefined by equivalence transformations, which do not change the result of the Penrose transform integral. Mathematically speaking, such quantities are representatives of cohomology classes, as identified in ref.~\cite{Eastwood:1981jy}. This then poses a puzzle, in that the non-linear product of twistor functions needed to obtain the Weyl double copy in position space is clearly inconsistent with the ability to first perform equivalence transformations. It thus seems that special representatives of each cohomology class must be chosen in each theory in order to make the double copy manifest, and it is not obvious a priori how to achieve this. A number of papers have subsequently addressed this point. First was ref.~\cite{Adamo:2021dfg}, which considered radiative spacetimes, and showed that data at past or future or null infinity could be used to pick out special cohomology representatives in twistor space. These were defined in terms of so-called {\it Dolbeault cohomology groups}, as distinct from the {\it \u{C}ech cohomology groups} that enter the original formulation of the Penrose transform. Reference~\cite{Chacon:2021lox} also considered the Dolbeault language, and argued that one can use known methods in Euclidean signature~\cite{Woodhouse:1985id} to pick out special twistor representatives for each spacetime field, so that a product structure is manifest in twistor space. It is not clear how the procedures of refs.~\cite{Adamo:2021dfg,Chacon:2021lox} are related, if at all.

More recently, ref.~\cite{Guevara:2021yud} considered the relationship between scattering amplitudes, twistor space, and classical solutions. It is known that certain classical solutions can be obtained as inverse Fourier transforms of momentum-space amplitudes. Reference~\cite{Guevara:2021yud} craftily split this inverse Fourier transform into two steps, where the first maps the amplitudes into quantities in twistor space. The second step then corresponds to the Penrose transform from twistor to position space, and it is therefore the case that scattering amplitudes themselves can be used to pick out cohomology representatives for certain classical fields in twistor space. As has been made clear elsewhere~\cite{Luna:2022dxo}, the representatives picked out by the relevant amplitudes in gauge, gravity or scalar theory are precisely those entering the original twistor double copy of refs.~\cite{White:2020sfn,Chacon:2021wbr}. Not only does this fix the cohomological ambiguities in the twistorial approach, it also establishes a very strong link between the double copies in momentum, twistor and position space.

Equally important as explaining the origin of known double copies is to continue to generalise this correspondence to novel theories or situations. In this spirit, refs.~\cite{Gonzalez:2022otg,Emond:2022uaf} recently proposed a new exact classical double copy for topologically massive gauge theory and gravity, called the {\it Cotton double copy}. Like the Weyl double copy in four dimensions, it uses the spinorial formalism of field theory, and expresses a precise relationship between scalar, gauge and gravity fields in position space. Indeed, this relationship is analogous to its four-dimensional counterpart, although appears to hold for a more restricted class of solutions than in the Weyl case. That is, both refs.~\cite{Gonzalez:2022otg,Emond:2022uaf} only found Cotton double copy examples in position space of Petrov type N, rather than the more general type D, and this fact demands a further explanation. It is also natural to ask whether there is a twistorial justification for the Cotton double copy, that mirrors its four-dimensional counterpart. Constructing such an argument should itself settle the issue of how general the Cotton double copy is, and this paper will show that this is indeed possible. We will use the language of {\it minitwistors} in three spacetime dimensions, and the presence of a topological mass means that we will have to consider an alternative to the usual Penrose transform. Just such a transform has been provided before in the mathematical literature~\cite{tsai_1996}. It is formulated by considering the most general possible cohomology classes in minitwistor space. This involves introducing an extra parameter in twistor space relative to the conventional four-dimensional case, whose presence corresponds to the presence of the topological mass in spacetime. Armed with this minitwistor transform, we will show explicitly that appropriately combining particular minitwistor representatives allows us to derive the position-space Cotton double copy.  

Our minitwistor derivation of the Cotton double copy will suffer from similar conceptual issues to its four-dimensional counterpart. Namely, the form of the double copy in minitwistor space involves products of ``functions", which should properly be interpreted as 
representatives of cohomology classes, with an appropriate procedure for picking them. However, the ideas of ref.~\cite{Guevara:2021yud} will once again come to the rescue: we will show that they can be generalised to the three-dimensional case, such that the minitwistor double copy follows as a consequence of the known double copy for scattering amplitudes in topologically massive theories. We will explicitly consider amplitudes corresponding to point-like sources emitting gauge bosons, which correspond to type D classical solutions\footnote{Strictly speaking we only consider linearized solutions, not exact solutions which are usually described within the Petrov classification.}. These will allow us to independently validate the form of the massive Penrose transform, where particular cohomology representatives are necessarily picked out. Interestingly, we will find that a simple twistor-space double copy occurs even in the type D case. However, a simple position-space double copy is restricted to type N only, as a direct consequence of the form of the massive Penrose transform. Our results provide a firm foundation for the Cotton double copy, whilst also providing an interesting counterpoint for the four-dimensional twistor double copy. This in turn suggests the use of twistor methods more widely in the study of (non-)exact classical double copies, including in higher dimensions where applicable.

The structure of our paper is as follows. In section~\ref{sec:review}, we review relevant properties of topologically massive theories in three dimensions, including the Cotton double copy. We also introduce the concept of minitwistors, and their associated Penrose transform, following ref.~\cite{tsai_1996}. In section~\ref{sec:derive}, we provide a twistorial derivation of the Cotton double copy, emphasising the similarities and differences with the four-dimensional twistor double copy of refs.~\cite{White:2020sfn,Chacon:2021wbr}. In section~\ref{sec:cohomology}, we show how the ideas of ref.~\cite{Guevara:2021yud} can be adapted to three dimensions, and use anyon solutions to illustrate our general arguments. Finally, we discuss our results and conclude in section~\ref{sec:discuss}.

\section{Review of necessary concepts}
\label{sec:review}

In this section, we review salient material for the rest of the paper, both in order to set up our notation and conventions, and also to make the presentation relatively self-contained. We begin by introducing the spinorial formalism for field theories. 

\subsection{Spinors in (2+1) dimensions}

Our first encounter with relativistic field theories in four spacetime dimensions typically involves the use of 4-vectors and tensors. As is well-known, however, it is possible to recast all relevant field equations into an alternative language, namely that of 2-component spinors~\cite{Witten:1959zza,Penrose:1960eq} (see e.g. refs.~\cite{Penrose:1985bww,Penrose:1986ca,Stewart:1990uf} for pedagogical reviews). Similar ideas occur in (2+1) dimensions \cite{Milson:2012ry,castillo20033,doi:10.1063/1.1592611}, which we now briefly review. 

We will be concerned with (dual) spinors $\lambda^A$ ($\lambda_A$), whose indices $A\in\{0,1\}$ may be raised and lowered using the two-dimensional Levi-Civita symbol:
\begin{equation}
    \lambda_B=\epsilon_{BA}\lambda^A,\quad
    \lambda^B=\epsilon^{BA}\lambda_A,\quad 
    \epsilon_{AB} = \begin{pmatrix}
		0 & -1 \\
		1 & 0
	\end{pmatrix} = -\epsilon^{AB}.
	\label{raiselower}
\end{equation}
Note that, in contrast to the well-known four-dimensional case, only one type of spinor index $A$ occurs. This is because the Lorentz group in (2+1) dimensions is covered by a single SL(2,C) group. In four dimensions, on the other hand, the Lorentz group is covered by two distinct SL(2,C) groups, leading to the presence of spinors $\lambda_A$ and conjugate spinors $\pi_{A'}$, where the prime is used to differentiate which SL(2,C) group acts on which index.  

We can convert (2+1)-dimensional tensors into spinors (and vice versa) using the {\it Infeld-van der Waerden symbols}, whose explicit form depends upon the chosen basis in spinor space. It is in fact possible to choose them to be real, so that we will adopt the $SL(2,\R)$ representation
\begin{equation}\label{paulibasis}
	\sigma^\mu_{AB} = \left\{\begin{pmatrix}
		1 & 0 \\
		0 & 1
	\end{pmatrix},\begin{pmatrix}
		0 & 1 \\
		1 & 0
	\end{pmatrix},\begin{pmatrix}
		1 & 0 \\
		0 & -1
	\end{pmatrix}\right\}.
\end{equation}
A given tensor index is thus converted into a {\it pair} of spinor indices, and one may also verify the following useful identities:
\begin{equation}
	\eta^{\mu\nu} = \frac12\sigma^{\mu}_{AB}\sigma^{\nu AB},~~~~~\sigma^\mu_{AB}\sigma_{\mu GD} = -(\epsilon_{AG}\epsilon_{BD}+\epsilon_{AD}\epsilon_{BG}). \label{eq:SigmaId}
\end{equation} 
As an example, a single 4-vector with real components has the spinorial translation
\[
p_{AB} =p_\mu\sigma^\mu_{AB} = \begin{pmatrix}
p_0 + p_2	&  p_1\\
p_1	& p_0 - p_2
\end{pmatrix}, ~~~~~\det(p_\mu\sigma^\mu_{AB}) = -p_\mu p^\mu,
\]
where the matrix thus obtained is referred to as a {\it bispinor}.
As may be verified by direct computation, one can always decompose a bispinor in (2+1) dimensions into the outer product of two complex spinors
\[
p_{AB} = \lambda_{(A}\bar{\lambda}_{B)},
\label{pAB}
\]
where the latter are given by
\[
\lambda_A = \frac{1}{\sqrt{p_2-p_0}}\begin{pmatrix}
	p_2-p_0 \\ p_1-im
\end{pmatrix},~~~~~\bar{\lambda}_A = \frac{1}{\sqrt{p_2-p_0}}\begin{pmatrix}
p_2-p_0 \\ p_1+im
\end{pmatrix}.
\]
In the scattering amplitudes literature, it is common to introduce a Dirac notation for (dual) spinors:
\begin{equation}
    \ket{\lambda}\equiv \lambda_A,\quad \bra{\lambda}\equiv \lambda^A.
    \label{braketdef}
\end{equation}
Then we can define spinor helicity variables in the usual way as
\[
\braket{\lambda^i\lambda^j} \equiv \epsilon^{AB}\lambda^i_B\lambda^j_A,
\]
where we note in particular the identities
\[
\braket{\bar{\lambda}\lambda} = 2im,~~~~~\braket{\lambda|\gamma^\mu|\bar{\lambda}} = -2p^\mu.
\]

\subsection{Topologically massive theories and their double copy}
Having reviewed the language of two-spinors in (2+1) dimensions, let us now introduce the theories that we will encounter throughout the paper. First up is {\it topologically massive Yang-Mills theory}, which is described by the action
\begin{equation}
    S_{TMYM}=\int d^3x\Bigg(-\frac{1}{4}F^{a\mu\nu}F_{a\mu\nu}+\epsilon_{\mu\nu\rho}\frac{m}{12}\left(6 A^{a\mu }\partial^{\nu}A^{\rho}_{a}+g\sqrt{2}f_{abc}A^{a\mu}A^{b\nu}A^{c\rho}\right)\Bigg),
\end{equation}
implying the equation of motion
\begin{equation}
	D_\mu F^{\mu\nu}+\frac{m}{2}\varepsilon^{\nu\rho \gamma}F_{\rho\gamma}=0 \ . \label{TMYM}
\end{equation}
Physically, this describes a gauge boson with mass $m$ and a single helicity $h = \frac{m}{|m|}$. The mass term in the action is only possible in three spacetime dimensions, due to the presence of the three-dimensional Levi-Civita tensor. Furthermore, unlike conventional mass terms in arbitrary spacetime dimension, one may show that the mass term introduced here is manifestly gauge-invariant. The mass is topological in the sense that it is independent of the local metric. 

We will also be concerned with {\it topologically massive gravity}, whose action is 
\begin{equation}
    S_{TMG}=\frac{1}{\kappa^2}\int d^3x\sqrt{-g}\left(-R-\frac{1}{2m}\epsilon^{\mu\nu\rho}\left(\Gamma^{\alpha}_{\mu\sigma}\partial_{\nu}\Gamma^{\sigma}_{\alpha\rho}+\frac{2}{3}\Gamma^{\alpha}_{\mu\sigma}\Gamma^{\sigma}_{\nu\beta}\Gamma^{\beta}_{\rho\alpha}\right)\right) \ ,
    \label{STMG}
\end{equation}
and leads to the equation of motion
\begin{equation}
    G_{\mu\nu}+\frac{1}{m}C_{\mu\nu}=0 \ ,
\end{equation}
where $C_{\mu\nu}$ is a symmetric tensor known as the {\it Cotton tensor}: 
\[
C_{\mu\nu} = \epsilon_{\mu\rho\sigma}D^\rho\left(R^{\sigma}_\nu - \frac{1}{4}\delta^\sigma_\nu R\right) \ . 
\]
We can think of this as a $(2+1)$-dimensional analogue of the Weyl tensor in four (or higher) dimensions, where the latter is what the Riemann curvature reduces to in the case of vacuum solutions of the Einstein equations. Like the Weyl tensor, the Cotton tensor vanishes for conformally flat spacetimes. As for the Yang-Mills case discussed above, the second term in the action of eq.~(\ref{STMG}) is impossible to write down in four spacetime dimensions. It is a correction to the pure Einstein-Hilbert action, and generates a mass for the graviton that is invariant under diffeomorphisms. 

As discussed in the introduction, it is by now very well-known that Yang-Mills theory and gravity (plus their generalisations) are related by the double copy, which applies to both scattering amplitudes and classical solutions. It was recently also conjectured that the topologically massive gauge and gravity theories considered here are related by a similar double copy \cite{Moynihan:2020ejh}, evidence for which has been presented in a number of non-trivial scenarios \cite{Burger:2021wss,Gonzalez:2021bes,Moynihan:2021rwh,Emond:2021lfy,Hang:2021oso,Gonzalez:2021ztm}. Important for this paper is the {\it Cotton double copy}~\cite{Gonzalez:2022otg,Emond:2022uaf}\footnote{Note that the Cotton tensor has appeared in a different off-shell double copy construction in \cite{Ben-Shahar:2021zww}.}, which directly relates classical solutions of the above equations of motion, expressed in the spinorial formalism. As reviewed in refs.~\cite{Gonzalez:2022otg,Emond:2022uaf}, the spinorial translation of the free field equation for Abelian topologically massive gauge theory is
\begin{equation}
{\pd^G}_A\Phi_{GB} = m\Phi_{AB},
\label{YMspinor}
\end{equation}
where $\partial_{AB}$ is the spinorial translation of the partial derivative operator $\partial_\mu$, and we have defined\footnote{In order to verify eq.~(\ref{YMspinor}), one must also use the relation $\epsilon^{\alpha \beta}\left(\partial_{\beta}^{\gamma} \varphi_{\gamma \alpha}-\partial_{\alpha}^{\gamma} \varphi_{\gamma \beta}\right)=0$, which follows from the well-known {\it Bianchi identity} for the field strength tensor.}
\[
\Phi_{AB} = \frac12\sigma^\mu_{AB}\epsilon_{\mu\nu\rho}F^{\nu\rho}.
\]
Similarly, the free-field equations of topologically massive gravity take the form
\begin{equation}
{\pd^E}_A C_{EBGD} = mC_{ABGD},
\label{Gspinor}
\end{equation}
where
\[
C_{ABGD} = C_{\mu\nu}\sigma^\mu_{AB}\sigma^\nu_{GD}.
\label{Cottonspinor}
\]
It is instructive to contrast these equations with their natural counterparts in four-dimensional gauge and gravity theory, namely the {\it massless free field equation}
\begin{equation}
    {\partial}^{A_1A'}\Phi_{A_1A_2\ldots A_{2n}}=0.
    \label{masslessfreefield}
\end{equation}
Here $\Phi_{A_1\ldots A_{2n}}$ is a multi-index symmetric spinor corresponding to a single polarisation state of the field\footnote{The other polarisation state obeys a similar equation to eq.~(\ref{masslessfreefield}), but with (un)primed indices interchanged.}, and the derivative operator is now the appropriate four-dimensional spinorial translation of the partial derivative operator in spacetime. We have written eq.~(\ref{masslessfreefield}) for a general spin-$n$, from which we may note that there are $2n$ spinor indices for a spin-$n$ field. Apart from the slight difference in derivatives, we see that eqs.~(\ref{YMspinor}, \ref{Gspinor}) differ from eq.~(\ref{masslessfreefield}) due to the presence of the mass term on the right-hand side. 

Solutions of eq.~(\ref{masslessfreefield}) of different spin can be related to each other by the {\it Weyl double copy}~\cite{Luna:2018dpt}, which has been shown to work for certain algebraically special spacetimes. 
\begin{equation}
    \Phi_{ABCD}=\frac{\Phi_{(AB}\Phi_{CD)}}{\Phi}.
    \label{WeylDC}
\end{equation}
Here $\Phi$ is a field satisfying the massless Klein-Gordon equation in spacetime, and to clarify where this applies, we may note that a consequence of the limited range of spinor indices is that an arbitrary multi-index symmetric spinor can be decomposed in terms of single-index {\it principal spinors}, such that we have
\begin{equation}
    \Phi_{ABCD}=\alpha_{(A}\beta_B\gamma_C\delta_{D)}.
    \label{principal}
\end{equation}
So-called {\it Petrov type D} solutions are those for which there are two distinct principal spinors, each of double multiplicity. Type N solutions have a single principal spinor of multiplicity four. These are the two cases of algebraically special solutions for which the Weyl double copy is known to be exact.

Motivated by the Weyl double copy, refs.~\cite{Gonzalez:2022otg,Emond:2022uaf} considered whether a similar relation can be written for topologically massive Yang-Mills and gravity theories. Indeed it can, provided one replaces the Weyl tensor with the Cotton tensor, and instead considers $\Phi$ to be a solution of the massive Klein-Gordon equation:
\begin{equation}
    C_{ABGD}=\frac{\Phi_{(AB}\Phi_{GD)}}{\Phi} \ . \label{eq:CottonDC}
\end{equation}
This is the {\it Cotton double copy} formula alluded to above, and is known to apply at least for type N solutions. In order to see whether it is in fact more general than this (as is the Weyl double copy), it is fruitful to seek a more underlying explanation of where the Cotton double copy comes from. In the case of the Weyl double copy, refs.~\cite{White:2020sfn,Chacon:2021wbr} provided a derivation of the position-space formula using the techniques of twistor theory. This  suggests that similar techniques could prove useful in deriving the Cotton double copy.  Before we can do this, however, we must first familiarise ourselves with twistor techniques in (2+1) dimensions. This is the subject of the following section.

\subsection{Minitwistor theory}

In this section, we give a brief introduction to the subject of twistors in three-dimensional space. Pedagogical reviews of four-dimensional twistor theory can be found in e.g. refs.~\cite{Penrose:1986ca,Huggett:1986fs,Adamo:2017qyl}. The subject of three-dimensional twistor theory is less well-known, and thus our aim is to collect a number of useful results from the literature in one place~\cite{Ward:1989vja,tsai_1996}. The relevant concepts are similar to the case of the four-dimensional twistor double copy defined in refs.~\cite{White:2020sfn,Chacon:2021wbr}: given flat spacetime, one may construct an abstract {\it twistor space}, such that points in spacetime are mapped non-locally to the latter and vice versa. Solutions of the massless free field equation of eq.~(\ref{masslessfreefield}) can be obtained as a certain contour integral in twistor space, which is known as the {\it Penrose transform}. In order to apply these same ideas to (2+1) dimensions,  we must first define the relevant twistor space, and then arrive at the necessary Penrose transform, which must somehow take into account the presence of the mass in topological gauge theory or gravity. Let us take each of these topics in turn.

\subsubsection{Minitwistor geometry}

\begin{figure}
    \centering
    \scalebox{0.5}{\includegraphics{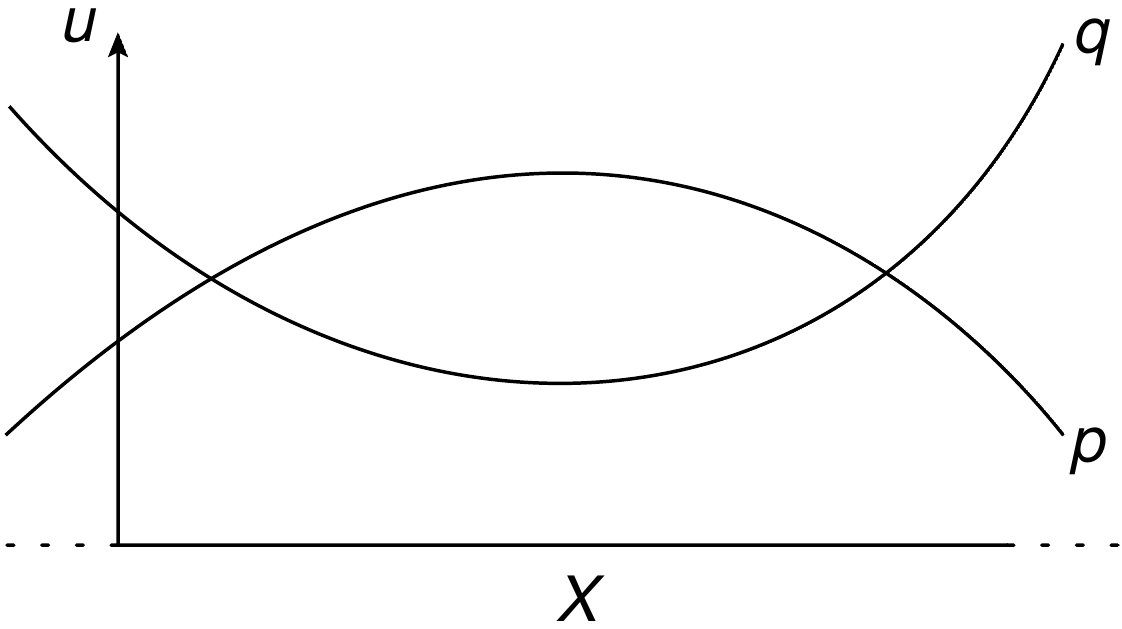}}
    \caption{Points $p$ and $q$ in spacetime can be visualised as curves in minitwistor space, where the coordinate $u$ is defined at each point on the Riemann spheres $X_p$ and $X_q$ corresponding to $p$ and $q$. Curves associated with different spacetime points can intersect in at least two places.}
    \label{fig:twistorcurve}
\end{figure}

Let us first consider complexified Minkowski space ${\cal M}=\mathbb{C}^3$, with line element
\begin{equation}
    ds^2=-dt^2+dx^2+dy^2,\quad t,x,y\in\mathbb{C}.
    \label{ds2}
\end{equation}
Using the Infeld-van-der-Waerden symbols of eq.~(\ref{paulibasis}), a point $x\in{\cal M}$ has a spinorial translation as a symmetric $2\times 2$ matrix:
\begin{equation}
    x^{AB}=\left(\begin{array}{cc}-t-y & -x\\ -x & -t+y
    \end{array}\right).
    \label{xAB}
\end{equation}
We may then define {\it minitwistor space} $\mathbb{MT}$ as the two-dimensional set of null planes in ${\cal M}$. Any such plane is defined by a null three-dimensional normal vector $n^\mu$, such that 
\begin{equation}
    n_\mu x^\mu=u,\quad n^2=0,\quad u\in\mathbb{C}.
    \label{nprops}
\end{equation}
Nullity of $n_\mu$, and the fact that it is defined only up to arbitrary scalings,  implies that its spinorial translation factorises as follows:
\begin{equation}
    n_{AB}\equiv n_\mu \sigma^\mu_{AB}=\lambda_A\lambda_B,
    \label{nnull}
\end{equation}
where $\lambda_A$ is itself only defined up to an overall complex scale:
\begin{equation}
    n^\mu\rightarrow \alpha^2 n^\mu\quad\Rightarrow\quad
    \lambda_A\rightarrow \alpha\lambda_A,\quad\alpha\in\mathbb{C}.
    \label{nrescale}
\end{equation}
The first condition in eq.~(\ref{nprops}) then implies
\begin{equation}
    u=x^{AB}\lambda_A\lambda_B \ , \label{eq:incidence}
\end{equation}
such that a given point in minitwistor space (representing a particular null plane) is described by coordinates
\begin{equation}
    Z^\alpha=(u,\lambda_A) \ ,
\label{ZAdef}
\end{equation}
satisfying the {\it incidence relation} of eq.~(\ref{eq:incidence}).
Equations~(\ref{nprops}, \ref{nrescale}) then imply that the coordinates appearing in eq.~(\ref{ZAdef}) are defined only up to the scalings 
\begin{equation}
   \left(u, \lambda_{A}\right) \sim\left(r^{2} u, r \lambda_{A}\right) \ , \label{eq:scaling}
\end{equation}
for $r \in \mathbb{C}^*$\footnote{Here and in what follows, $\mathbb{C}^*$ denotes the set of non-zero complex numbers.}. A spinor $\lambda_A$ has two complex components, which reduces to one if an overall complex scale is removed. Thus, $\lambda_A$ defines a point on the Riemann sphere $\mathbb{CP}^1$. In general, we need two coordinate patches to cover the sphere, which we may choose as
\begin{align}
    &U_0:\quad \lambda_A=(1,z)\\
    &U_1:\quad \lambda_A=(w,1).
    \label{U0U1}
\end{align}
Given that $\lambda_A$ is defined only up to rescalings, we may identify $z=w^{-1}$ on the overlap $U_0\cap U_1$. The single complex coordinate $u$ is defined at each point on the Riemann sphere, and thus we may think of $\mathbb{MT}$ as a fibre bundle, with $\mathbb{CP}^1$ as the base space.
Formally speaking, it is the holomorphic tangent bundle $T\mathbb{CP}^1$ of the Riemann sphere. In particular, a given point in minitwistor space assigns a holomorphic tangent vector to each point on the Riemann sphere associated with $\lambda_A$. To see this, note that a general holomorphic vector field on $\mathbb{CP}^1$ may be written as
\begin{equation}
    f(z)\partial_z =\left(\sum_{n=0}^\infty a_n z^n\right)\partial_z
    =-\left(\sum_{n=0}^\infty a_n w^{2-n}\right)\partial_w,
    \label{vechol}
\end{equation}
where we have used $\partial_z=-w^2\partial_w$ in the second equality. Holomorphicity in both coordinate choices (in particular the absence of poles) then implies $a_n=0$ for $n>2$, such that a general holomorphic vector field on $U_0\cap U_1$ may be written as 
\begin{equation}
    (a_0+a_1z +a_2 z^2)\partial_z.
    \label{vecquad}
\end{equation}
The incidence relation of eq.~(\ref{eq:incidence}) can be expanded in $U_0$ using eq.~(\ref{xAB}) as 
\begin{equation}
    u=(-t+y)z^2-2xz-(y+t),
    \label{incidencez}
\end{equation}
From eq.~(\ref{vecquad}), this defines a holomorphic vector field 
\begin{displaymath}
u(z)\partial_z
\end{displaymath}
on $\mathbb{CP}^1$, as required\footnote{To fully identify $\mathbb{MT}$ with the holomorphic tangent bundle of $\mathbb{CP}^1$, one must show that {\it all} possible holomorphic vector fields can be obtained from the incidence relation. We will not formally prove this here.}. As described above and as is hopefully clear from eq.~(\ref{incidencez}), a fixed spacetime point $x$ and picks out a specific vector at each point on the Riemann sphere associated with $\lambda$. Thus, the coordinate $u$ defines a section 
\begin{equation}
    u:\quad \mathbb{CP}^1 \rightarrow T\mathbb{CP}^1
\end{equation}
of the holomorphic tangent bundle of the Riemann sphere. Thus, a point in spacetime corresponds to a section of the holomorphic tangent bundle of a Riemann sphere $X$. As explained in ref.~\cite{Ward:1989vja}, we may visualise this in twistor space as shown in figure~\ref{fig:twistorcurve}. The horizontal axis shows the Riemann sphere $X$ represented as a complex line. The vertical axis then denotes the value of $u$ at each point on $X$, such that we may visualise this as a curve. 

So far we have worked in complexified Minkowski space. If we wish to use real coordinates in Lorentzian signature, then the matrix given in eq.~(\ref{xAB}) will be real; that is, we choose to impose the reality condition $(u,\lambda_A)\sim(\bar{u},\bar{\lambda}_A)$. Now, the incidence relation of eq.~(\ref{eq:incidence}) defines a real null-plane or a timelike line, depending on whether $u$ and $\lambda_0/\lambda_1$ are (non-)real.

\subsubsection{Dimensional reduction}

In what follows, we will have to obtain the relevant Penrose transform that converts data in minitwistor space into solutions of the topological gauge and gravity equations in spacetime. In doing so, we will find it useful to rely on some alternative ways of thinking about $\mathbb{MT}$. The first of these relies on the more well-known concept of twistor space for four-dimensional complexified Minkowski spacetime ${\cal M}_4=\mathbb{C}^4$. As remarked already above, we must consider two types of spinor in four spacetime dimensions, given that the Lorentz group is covered by two distinct SL(2,$\mathbb{C})$ groups, which we may refer to as SL(2,$\mathbb{C})_L$ and SL(2,$\mathbb{C})_R$. These act on (conjugate) spinors, which carry (un-)primed indices respectively. To convert a given tensor or 4-vector into the spinorial language, we can again contract with the relevant Infeld-van-der-Waerden symbols, for which a suitable choice is
\begin{equation}
	\sigma^\mu_{AA'} = \left\{\begin{pmatrix}
		1 & 0 \\
		0 & 1
	\end{pmatrix},\begin{pmatrix}
		0 & 1 \\
		1 & 0
	\end{pmatrix},\begin{pmatrix}
		1 & 0 \\
		0 & -1
	\end{pmatrix},
	\begin{pmatrix}
		0 & -i \\
		i & 0
	\end{pmatrix}
	\right\}.
\label{Infeld4}
\end{equation}
Comparing with eq.~(\ref{paulibasis}), we see that the Infeld-van-der-Waerden symbols carry a spinor index associated with each of the groups SL$(2,\mathbb{C})_{L,R}$, and we have also appended the "missing" Pauli matrix to be the third component $\sigma^3_{AA'}$, where now $\mu\in\{0,1,2,3\}$. With these conventions, one has
\begin{equation}
   x_{AA'}=\left(\begin{array}{cc} -t+y & x -iz\\ x+iz & -t-y
   \end{array}\right),\quad
   x^{AA'}=\left(\begin{array}{cc} -t-y & -x -iz\\ -x+iz & y-t
   \end{array}\right).
    \label{XAA'}
\end{equation}
The twistor space corresponding to four-dimensional Minkowski spacetime can also be identified with the space of certain null 2-planes. It turns out that these can be parameterised by twistor coordinates\footnote{For want of a better notation, we will use (non-)calligraphic symbols to refer to (three-) four-dimensional twistors respectively.}
\begin{equation}
    {\cal Z}^\alpha=(\mu^{A'},\lambda_A),
    \label{4dtwistor}
\end{equation}
subject to the incidence relation
\begin{equation}
    \mu^{A'}=x^{AA'}\lambda_A.
    \label{incidence4}
\end{equation}
Twistors satisfying this relation are defined only up to an overall rescaling
\begin{equation}
    {\cal Z}^\alpha\sim r{\cal Z}^\alpha,\quad \Rightarrow\quad
    (\mu^{A'},\lambda_A)\sim (r\mu^{A'},r\lambda_A),
    \label{Zrescale}
\end{equation}
and are said to live in {\it projective twistor space} $\mathbb{PT}$.
We may now obtain minitwistor space of $\mathbb{C}^3$ by {\it dimensionally reducing} four-dimensional twistor space. After dimensional reduction, one may choose to work with real coordinates in a specific signature by picking a reality condition as in the previous section. To see how this works note that, in our above examples, we can isolate the three-dimensional coordinates from $x^{AA'}$ by introducing a constant vector with spinorial translation  
\begin{equation}
T_{BA'}=\left(\begin{array}{cc} 0& 1 \\ -1 & 0
\end{array}\right),
\label{Tdef}
\end{equation}
and forming the combination
\begin{equation} \label{eq:xT2}
x_{AA'}{T_B}^{A'}=x_{AA'}\epsilon^{A'C'}T_{BC'}=
x_{AB}+\left(\begin{array}{cc}0&-iz\\iz&0\end{array}\right),
\end{equation}
as follows from explicit computation. Then we may write
\begin{equation}
    x^{AB}={x^{(A}}_{A'}T^{B)A'},
    \label{xAB43}
\end{equation}
whose geometric interpretation is that we are ignoring translations along the direction of $T^\mu$ in spacetime. Furthermore, SL(2,$\mathbb{C}$) covariance of eq.~(\ref{xAB43}) means that we can pick {\it any} direction in spacetime in order to perform the dimensional reduction. Removing the symmetrising brackets in eq.~(\ref{xAB43}) will generate an antisymmetric contribution on the left-hand side, which for a two-dimensional matrix must be proportional to the Levi-Civita symbol. Thus, on general grounds we may write~\cite{Hitchin:1982gh,Jones_1985,Adamo:2017xaf}
\begin{equation} \label{eq:xT}
{x^{A}}_{A'} T^{BA'}=x^{AB}+ b \epsilon^{AB} \ ,
\end{equation}
where taking the determinant of both sides can be used to infer the relation
\begin{equation}
   b=i\sqrt{x_{\text{4d}}^2-x_{\text{3d}}^2}.
   \label{bresult}
\end{equation}
Using eq.~(\ref{xAB43}), we can recover the minitwistor incidence relation in eq.~\eqref{eq:incidence} from the 4d incidence relation of eq.~(\ref{incidence4}). To do so, one may define
\begin{equation}
u\equiv \mu_{A'}T^{AA'}\lambda_A \ .
\label{u4def}
\end{equation}
This can be shown to be invariant under the equivalent of translations along the vector field $T^\mu$ in twistor space~\cite{Adamo:2017xaf}. Also, the four-dimensional twistor scaling property of eq.~(\ref{Zrescale}) implies $u\rightarrow r^2 u$, as required in eq.~(\ref{eq:scaling}). 
Combining eqs.~(\ref{incidence4}, \ref{eq:xT}, \ref{u4def}), we find
\begin{equation}
    u=(x^{AB}\lambda_A+b\lambda^B)\lambda_B,
    \label{u4def2}
\end{equation}
where the second term in the brackets corresponds to the effect of a translation in the $T^\mu$ direction. This vanishes after contracting with the spinor $\lambda_B$ outside the brackets, but suggests that the most general equivalence relation in the three-dimensional twistor coordinates is
\begin{equation}
        (x^{AB} \lambda_{A}\lambda_{B},\lambda_A) \sim (r^2 (x^{AB} \lambda_{A}+b\lambda^B)\lambda_{B},r \lambda_A) .
\end{equation}
Indeed, this motivates another way to define coordinates on minitwistor space, as
\begin{equation} \label{eq:Z two spinors}
Z^A=(\mu^{A},\lambda_A)\ , \quad \mu^{A}=x^{AB} \lambda_{B} \ ,
\end{equation}
which arise more naturally from the dimensional reduction point of view. In this case, the coordinates are defined up to the following equivalence:
\begin{equation}
(\mu^{A},\lambda_A) \sim (r (\mu^{A}+b\lambda^A),r \lambda_A) \ . \label{eq:FullEquiv}
\end{equation}
In discussing the Penrose transform we need for topological gauge and gravity theory, it is useful to discuss yet another, and rather more formal, way to describe minitwistor space. Recall that $\bb{MT}$ consists of the set of null two-planes in $\bb{C}^3$. Reference~\cite{tsai_1996} considers Euclidean signature for the latter, and points out that the complex Euclidean group of transformations that define the space (rotations plus translations) is covered by the group ESL(2,$\bb{C}$). If we then quotient this group by the group of isometries of null planes, we will obtain the group that acts on minitwistor space. The elements of the former group can be written as
\begin{equation*}
\mathrm{ESL}(2,\mathbb{C})=\{(A,B)|A\in \mathrm{SL}(2,\mathbb{C}), B \in \{\text{2$\times$2 complex trace-free matrices}\}\},
\end{equation*}
subject to the composition law
\begin{equation}
    (A,B)\circ (A',B')=(AA',AB'A^{-1}+B).
    \label{ABcompose}
\end{equation}
We thus see that $A$ is an SL(2,$\bb{C}$) transformation associated with rotations, and $B$ is associated with translations. The relevant closed subgroup that we must quotient out is given by~\cite{tsai_1996}
\begin{equation}
Q=\left\{\left(R=\left(\begin{array}{cc}
r & t \\
0 & r^{-1}
\end{array}\right),S=\left(\begin{array}{cc}
b & a \\
0 & -b
\end{array}\right)\right) \ ; \  a, b, r, t \in \mathbb{C}, r \neq 0\right\} \ , \label{eq:quotient}
\end{equation}
and to show that this is correct, we can simply apply the equivalence relation $g\sim gq$ (with $g \in \mathrm{ESL}(2,\mathbb{C})$, $q \in Q$), and show that this corresponds to the equivalence relation of eq.~(\ref{eq:FullEquiv}) when acting on minitwistor coordinates. One may then parametrise
\begin{equation}
 g=(A_{AB}=(\lambda_A \,\chi_B)\ ,\ X=x^{AB}) \ , 
 \end{equation}
 where $\chi_B$ is an arbitrary spinor. From eq.~(\ref{ABcompose}), $Q$ then acts on $g$ as
 \begin{equation}
  gq=(AR\ ,\ X+ASA^{-1})   \ .  \label{eq:actionQonESL}
 \end{equation}
Expanding appropriately and using eq.~\eqref{eq:quotient}, one obtains the correspondence~\cite{tsai_1996}
\begin{equation}
 g\rightarrow g q: \quad \lambda_A\rightarrow r  \lambda_A \ , \quad x^{AB}\lambda_A\rightarrow r  (x^{AB}\lambda_A + b \lambda_A) \ . \label{eq:actionQonMT}
\end{equation}
We can then interpret $\lambda_A$ and $\mu^B=x^{AB}\lambda_A$ as coordinates on minitwistor space, subject to equivalence relations which indeed match those found by dimensional reduction in eq.~(\ref{eq:FullEquiv}).

\subsubsection{The Penrose transform for massless free fields}

Having introduced minitwistor space in various ways, our next task is to find the appropriate Penrose transform that expresses spacetime fields as contour integrals in minitwistor space. To this end, let us first recall the Penrose transform in four-dimensional twistor theory~\cite{Penrose:1969ae}. Solutions of the four-dimensional massless free field equation of eq.~(\ref{masslessfreefield}) can be expressed via the following contour integral:
\begin{equation}
    \phi_{A_1A_2\ldots A_{2n}}=\frac{1}{2\pi i}\oint_\Gamma
    \braket{\lambda d\lambda}
    \lambda_{A_1}\lambda_{A_2}\ldots \lambda_{A_{2n}}\rho_x[f({\cal Z}^\alpha)],\quad \braket{\lambda d\lambda}=\lambda_E d\lambda^E.
    \label{Penrose}
\end{equation}
Here the contour $\Gamma$ lies on the Riemann sphere $X$ associated with a given spacetime point $x$, and $\lambda_A$ is the spinor that forms half of the twistor components of eq.~(\ref{4dtwistor}). There is then a holomorphic function $f({\cal Z}^\alpha)$ of twistor coordinates, where the symbol $\rho_x$ denotes restriction to the Riemann sphere $X$, such that all twistors obey the incidence relation of eq.~(\ref{incidence4}). The contour $\Gamma$ must be such that it separates any poles of $f({\cal Z}^\alpha)$, and for there to be a non-zero answer, there must be at least two poles, one on either side of $\Gamma$. We may take the latter to correspond to the equator of the Riemann sphere without loss of generality.

As is well-known~\cite{Eastwood:1981jy}, the ``functions" $f({\cal Z}^\alpha)$ are not unique, but can be subjected to equivalence transformations that do not affect the result of the contour integral:
\begin{equation}
    f({\cal Z}^\alpha)\sim f({\cal Z}^\alpha)+f_N({\cal Z}^\alpha)+f_S({\cal Z}^\alpha),
    \label{fequiv}
\end{equation}
where $f_N({\cal Z}^\alpha)$ ($f_S({\cal Z}^\alpha)$) has poles only in the northern (southern) hemisphere of $X$ respectively. Substituting eq.~(\ref{fequiv}) into eq.~(\ref{Penrose}), we may evaluate the additional contributions by simply closing the contour in the opposite side to where the poles are, giving rise to a zero result, as required. In more formal mathematical terms, we say that the quantities $f({\cal Z}^\alpha)$ are representatives of {\it (\u{C}ech) cohomology classes}, and a fuller exposition of this in the present context can be found in ref.~\cite{Chacon:2021lox}. There is, however, a further restriction on $f({\cal Z}^\alpha)$, arising from the fact that twistors obeying the incidence relation constitute points in projective twistor space $\mathbb{PT}$, and thus are only defined up to the rescalings of eq.~(\ref{Zrescale}). If the integrand and measure in eq.~(\ref{Penrose}) are to be invariant under ${\cal Z}^\alpha\rightarrow r{\cal Z}^\alpha$, then it must be the case that $f({\cal Z})$ is a homogeneous function of degree $(-2n-2)$, for a spin-$n$ spacetime field:
\begin{equation}
    f(r{\cal Z}^\alpha)=r^{-2n-2}f({\cal Z}^\alpha).
    \label{fndef}
\end{equation}
Denoting holomorphic functions on minitwistor space of homogeneity $N$ by ${\cal O}(N)$, we would then say in formal language that the Penrose transform is an isomorphism between spacetime fields of spin $n$, and elements of the \u{C}ech cohomology group\footnote{Strictly speaking, the Penrose transform of eq.~(\ref{Penrose}) relates to {\it sheaf cohomology groups}, where ${\cal O}(N)$ then denotes the sheaf of holomorphic functions of homogeneity $N$. However, \u{C}ech cohomology provides a suitable approximation to sheaf cohomology for all practical purposes here. See e.g. ref.~\cite{Huggett:1986fs} for a pedagogical discussion of this point.} $H^1(\mathbb{PT},{\cal O}(-2n-2))$. 

It is straightforward to write down a Penrose transform for solutions of the three-dimensional massless free field equation. The latter is given by
\begin{equation}
    \partial^{A_{1} B} \phi_{A_{1} \cdots A_{2n}}(x)=0 \ , \label{eq:massless}
\end{equation}
and the Penrose transform itself by~\cite{Ward:1989vja}
\begin{equation}
\phi_{A_1\ldots A_{2n}}=\frac{1}{2\pi i}\oint_\Gamma 
\braket{\lambda d\lambda}\lambda_{A_1}\ldots 
\lambda_{A_{2n}} \rho_x[f(Z^\alpha)].
    \label{Penrose2}
\end{equation}
This is directly analogous to eq.~(\ref{Penrose}), where $\Gamma$ is again a contour on the Riemann sphere $X$ associated with the spacetime point $x$, and $f(Z^\alpha)$ a holomorphic function of the minitwistor coordinates of eq.~(\ref{ZAdef}). 

Above, we have presented the Penrose transforms for massless free fields in the language of \u{C}ech cohomology, in which they take the form of contour integrals in (mini-)twistor space. An alternative approach exists, in which twistor integrands are interpreted using differential forms, and the freedom to redefine twistor integrands is interpreted using {\it Dolbeault cohomology} (see e.g. ref.~\cite{Chacon:2021lox} for a recent detailed comparison of the two approaches). We will remain with the \u{C}ech approach in what follows, which will turn out to be more convenient for our purposes. However, an obvious deficiency of eq.~(\ref{Penrose2}) is that it only works for massless free fields, and thus is inapplicable to topologically massive gauge and gravity theory. That it is possible to generalise the three-dimensional Penrose transform to incorporate (topological) mass is possible on very general grounds, which we review in the following section.

\subsubsection{The Penrose transform for massive free fields}

A three-dimensional Penrose transform for massive fields has been presented in the twistor literature by Tsai~\cite{tsai_1996}, whose starting point is to consider the above construction of minitwistor space $\mathbb{MT}$ as the quotient space $G/H$, where $G$=ESL(2,$\mathbb{C})$ is the universal cover of the complex Euclidean group that generates all points in $\mathbb{C}^3$ and $H=Q$ is the group of isometries of null planes. To construct a Penrose transform, we must consider defining functions on $\mathbb{MT}$. Functions at a point form a vector space $V$ under addition, and we must therefore consider a mathematical structure consisting of a copy of $V$ associated with all points in $G/H$, such that one has a type of fibre bundle. In fact, this structure is known as a {\it homogeneous vector bundle}, where the word ``homogeneous" refers to the fact that the base space is itself a quotient space. There is a canonical way to construct homogeneous vector bundles (see e.g. ref.~\cite{Ward:1990vs}), as follows. We can first think of constructing a conventional vector bundle on $G$ by placing a copy of $V$ (the ``fibre") above each point of $G$, and then letting a representation $\rho(g)$ of each element $g\in G$ act on vectors $v\in V$. Then, given $g\in G$ and $v\in V$, we may identify points in this vector bundle by asserting the equivalence
\begin{equation}
    (g,v)\sim (gh,\rho(h^{-1}) v), h\in H.
    \label{gvequiv}
\end{equation}
The first component of this relation tells us that $gh$ is to be identified with $g$, which is simply the action of quotienting out $G$ by the closed subgroup $H$. The second component then implements the fact that the vectors in the fibres above $g$ and $gh$ must be identified\footnote{To see why $h^{-1}$ rather than $h$ occurs in the second relation in eq.~(\ref{gvequiv}), one may demand that the effect of acting on both components with a group element $h_1h_2\in H$ is the same as acting first with $h_2$, then with $h_1$. The group $H$ acts towards the left on elements of $G$, but towards the right on elements of $V$, such that the inverse ensures that the ordering of successive transformations acting on $V$ is correct.}. 

Returning to the specific case of $G$=ESL(2,$\mathbb{C})$ and $H=Q$, we will be considering scalar functions, which must then be acted on by one-dimensional representations of $Q$. From eqs.~(\ref{ABcompose}, \ref{eq:quotient}), one finds
\begin{equation}
    (R_1,S_1)\circ (R_2,S_2)=
    \left( \left(\begin{array}{cc} r_1 r_2 & \ast \\ 0 & (r_1 r_2)^{-1}
    \end{array}\right), \left(\begin{array}{cc}
    b_1+b_2 & \ast \\ 0 & -b_1-b_2\end{array}\right)\right),
    \label{Qcombine}
\end{equation}
and thus one sees that the $r$ parameters are multiplicative, whereas the $b$ parameters are additive. Physically, this is related to the fact that the former are associated with rotations, and the latter with translations. A one-dimensional representation that embodies these properties can be easily written down as
\begin{equation}
    \rho((R,S))=r^{-N} e^{Mb},\quad N\in \mathbb{Z},\quad M\in\mathbb{C}.
    \label{rep1}
\end{equation}
Indeed, this represents an infinite family of one-dimensional representations, one for each combination $(N,M)$, and our reason for restricting $N$ to be an integer will be clarified below. We have seen 
that, acting on minitwistor coordinates $(\mu^A,\lambda_A)$, transformations $h\in Q$ act according to eq.~(\ref{eq:FullEquiv}). Thus, functions acted on by the representation of eq.~(\ref{rep1}) must satisfy
\begin{equation} \label{eq:conditionTsai}
 \check{f}_M(r(\mu^A+b \lambda^A) ,r \lambda_A)=r^{N} e^{-M b} \check{f}_M(\mu^A , \lambda_A) \ .
\end{equation}
To make sense of this condition, we can consider the case $M=0$, for which there will no $b$ parameter in eq.~(\ref{rep1}). Then eq.~(\ref{eq:conditionTsai}) reduces to 
\begin{displaymath}
    \check{f}_0(r\mu^A,r \lambda_A)=r^{N} \check{f}_0(\mu^A , \lambda_A),
\end{displaymath}
which is merely the requirement that the function $f_0$ be homogeneous with degree $N$. Such functions enter the massless Penrose transforms in three and four dimensions, where $N=-2n-2$ is related to the spin $n$ of the spacetime field. For non-zero $M$, the  parameter $b$ in eq.(\ref{eq:conditionTsai}) corresponds to the additional freedom to redefine minitwistor coordinates, as in eq.~(\ref{eq:FullEquiv}).
Given this more general class of functions, we can construct a generalised Penrose transform. First, by setting $r=1$ in Eq.~\eqref{eq:conditionTsai}, differentiating this equation with respect to $b$, and evaluating it at $b=0$ we get that $\check{f}_M$ must obey:
\begin{equation}
\lambda_{A}\frac{\partial \check{f}_M(Z)}{\partial \mu_{A}}=-M \check{f}_M(Z),
\end{equation} 
A solution of this equation can be written as:
\begin{equation} \label{eq:solution f}
\check{f}_M(Z)=e^{-M\frac{\braket{a \mu}}{\braket{a \lambda}}} \check{g}(\lambda_A,\braket{\mu \lambda}).
\end{equation} 
where $a_A$ is an arbitrary spinor. To be compatible with \eqref{eq:conditionTsai} $\check{g}$ must be homogeneous of degree $N$, which we will choose as above to be $N=-2n-2$ (it is for this reason that we have chosen $N\in\mathbb{Z}$ in eq.~(\ref{rep1})). We may then consider the contour integral
\begin{equation}
\phi_{A_1\ldots A_{2n}}=\frac{1}{2\pi i}
\oint_\Gamma \braket{\lambda d\lambda}\lambda_{A_1}
\ldots \lambda_{A_{2n}}\rho_x[ \check{f}_M(Z^\alpha)],
\label{PenroseM}
\end{equation}
which consists of simply replacing the ``function" in the massless three-dimensional Penrose transform of eq.~(\ref{Penrose2}) with one of the more general types of function defined above. Acting on both sides with a derivative operator, we find
\begin{align}
{\nabla^{B}}_{A_1}\phi_{BA_2\ldots A_{2n}}(x)&=-\frac{1}{2\pi i}\oint_{\mathcal{C}}\langle\lambda d \lambda\rangle 2 \lambda_{B} \lambda_{A_2} \ldots \lambda_{A_{2n}} \lambda^{(B}\rho_x\left[\frac{\partial \check{f}_m(Z)}{\partial \mu^{A_1)}}\right]  \nonumber\\ 
&= \frac{M}{2\pi i} \oint_{\mathcal{C}}\langle\lambda d \lambda\rangle \lambda_{A_1} \lambda_{A_2} \ldots \lambda_{A_{2n}} \rho_x[\check{f}_M(Z)]  \nonumber \\
&= M \phi_{A_1\ldots A_{2n}}(x) \ ,
\label{massivefreefield}
\end{align}
where in the first line we have used
\begin{displaymath}
    {\nabla^{B}}_{A_1}f={(\sigma^\mu)^{B}}_{A_1}\sigma_\mu^{DC}\lambda_C\frac{\partial f}{\partial \mu^D},
\end{displaymath}
together with eq.~\eqref{eq:SigmaId}. This tells us two things: (i) $\phi_{A_1\ldots A_{2n}}$ constructed in this manner satisfies the massive free field equation of topologically massive gauge theory and gravity (eqs.~(\ref{YMspinor}, \ref{Gspinor})); (ii) the parameter $M$, which arose above in classifying the most general type of functions that can be defined on minitwistor space, can be identified with the topological mass $m$. We will thus make this identification in what follows. 

Note that a more general solution to eq.~\eqref{eq:conditionTsai} can be constructed as a sum over different arbitrary spinors $a_{A}$ and homogeneous functions $g_i$:
\begin{equation}\label{eq:f massive}
\check{f}_m(Z)=\sum_{a,i} e^{-m\frac{\braket{a \mu}}{\braket{a \lambda}}} \check{g}_i(\lambda_{\alpha},\braket{\mu \lambda}) \ .
\end{equation}
That this satisfies the general massive free field equation can be verified by explicit calculation, but anyway follows from linearity of the field equation.

As in the massless case, the twistor functions $\check{f}_m(Z)$ entering the Penrose transform of eq.~(\ref{PenroseM}) are not actually functions but defined only up to equivalence transformations, in this case of the form
\begin{equation}
\check{f}_m(Z)\sim\check{f}_m(Z)+e^{-m\frac{\braket{a_N \mu}}{\braket{a_N \lambda}}} \check{g}_N'(\lambda_{\alpha},\braket{\mu \lambda})
+e^{-m\frac{\braket{a_S \mu}}{\braket{a_S \lambda}}} \check{g}'_S(\lambda_{\alpha},\braket{\mu \lambda}) \ ,
\end{equation}
where $\check{g}'_{N,S}$ has poles only in the northern and southern hemispheres of $X$ respectively, and $\braket{a_{N(S)} \lambda}\neq 0$ in the southern (northern) hemisphere. In formal mathematical language, we would say that the quantity $\check{f}_M(Z^\alpha)$ is a representative of a cohomology class, which is itself a member of the \u{C}ech cohomology group $H^1(\mathbb{MT},{\cal O}(N,M))$, where ${\cal O}(N,M)$ denotes the functions acted on by eq.~(\ref{rep1}). Note that our above arguments merely show that a cohomology class in mini-twistor space gives a solution of the field equations in spacetime. However, ref.~\cite{tsai_1996} proves (see Proposition 2.10) that all possible solutions can be obtained in this way, so that the relationship between fields and cohomology classes is formally an isomorphism. 

In this section, we have reviewed a particular generalised Penrose transform on minitwistor space, whose ``functions" correspond to cohomology classes labelled by two parameters $(N,M)$. The first of these represents the homogeneity of the cohomology representative, and the second turns out to correspond to the mass in the field equations of topologically massive gauge theory and gravity in three spacetime dimensions. The latter are special cases of eq.~(\ref{massivefreefield}), but we will also need the spinless case, for which one may verify that the spacetime field $\phi$ satisfies the massive Klein-Gordon equation
\begin{equation}
    (\partial^2+m^2)\phi=0.
    \label{massiveKG}
\end{equation}
In four spacetime dimensions, the Penrose transform may be used to show that the position-space double copy for massless free fields can be derived from a certain procedure in twistor space~\cite{White:2020sfn,Chacon:2021wbr}. Now that we have identified the appropriate Penrose transform for minitwistor space, we can perform a similar analysis for topologically massive gauge and gravity theory in three spacetime dimensions.

\section{A minitwistor derivation of the Cotton double copy}
\label{sec:derive}

We have seen that the generalised Penrose transform of eq.~(\ref{PenroseM}) identifies solutions of the massive field equation of eq.~(\ref{massivefreefield}) with holomorphic twistor ``functions" (cohomology class representatives) having the form of eq.~(\ref{eq:solution f}). The homogeneity of a spin-$n$ field was found above to be $N=-2n-2$, and thus a scalar, gauge and gravity field will be associated with twistor representatives of homogeneity $-2$, $-4$ and $-6$ respectively. Let us introduce a scalar representative $\check{f}_{-2}(Z^\alpha)$, and a pair of gauge theory representatives $\check{f}^{(i)}_{-4}(Z^\alpha)$ ($i\in \{1,2\}$):
\begin{equation}
    \check{f}_{-2}(Z^\alpha)=e^{-m\frac{\braket{a\mu}}
    {\braket{a\lambda}}} \check{g}_{-2}(u,\lambda_A),\quad
    \check{f}_{-4}(Z^\alpha)=e^{-m\frac{\braket{a\mu}}
    {\braket{a\lambda}}} \check{g}_{-4}(u,\lambda_A) \ ,
    \label{twistorreps}
\end{equation}
where $\check{g}_{N}$ is an homogeneous ``function'' of degree $N$. It follows that one may construct a gravitational twistor representative by forming the product
\begin{equation}
    \check{f}_{-6}(Z^\alpha)=\frac{\check{f}^{(1)}_{-4}(Z^\alpha)
    \check{f}^{(2)}_{-4}(Z^\alpha)}{\check{f}(Z^\alpha)}=
    e^{-m\frac{\braket{a\mu}}
    {\braket{a\lambda}}}\check{g}_{-6}(u,\lambda_A),
    \label{eq:CechDC}
\end{equation}
with
\begin{equation}
      \check{g}_{-6}(u,\lambda_A)=\frac{\check{g}^{(1)}_{-4}(u,\lambda_A)
    \check{g}^{(2)}_{-4}(u,\lambda_A)}{\check{g}_{-2}(u,\lambda_A)}.
    \label{g-6}
\end{equation}
In the four-dimensional case of refs.~\cite{White:2020sfn,Chacon:2021wbr}, it was argued that choosing certain representatives allows to derive the Weyl double copy in position space. We may do something very similar here, in order to obtain the Cotton double copy. To see how this works, we may first recall that for twistor representatives with at most two poles on the Riemann sphere $X$, a $p-$fold pole gives rise to a $(2n-p+1)-$fold principal spinor of the corresponding spacetime field, at the point $x$ (see e.g. ref.~\cite{Penrose:1986ca}, and ref.~\cite{Chacon:2021wbr} for a more recent discussion of this point). We may then consider the representatives
\begin{equation}
    \check{f}_{-2-2n}(Z^\alpha)=e^{-m\frac{\braket{a \mu}}{\braket{a \lambda}}} \frac{G(u,\lambda_A)}{(\chi(u,\lambda_A))^p} \ . \label{eq:twistF}  
\end{equation}
Here $G(u,\lambda_A)$ and $\chi(u,\lambda_A)$ are homogeneous and holomorphic minitwistor functions, such that $\chi(u,\lambda_A)$ has $q\leq 2n$ simple zeros, corresponding to poles in $\check{f}_{-2-2n}(Z^\alpha)$ enclosed by the contour $\mathcal{C}$. Furthermore, $G(u,\lambda_A)$ is regular at the $p$-fold pole given by the zero of $\chi(u,\lambda_A)$. For Type N solutions in which the field has only one $2n$-fold principal spinor, $\chi(u,\lambda_A)$ has a simple zero and $p=1$. For Type D solutions which have 2 different $n$-fold principal spinors, $\chi(u,\lambda_A)$ has two simple zeros and $p=n+1$. In refs.~\cite{Gonzalez:2022otg,Emond:2022uaf}, the Cotton double copy was explicitly argued to hold in position space for type N solutions only. Thus, we will shortly show how the type N Cotton double copy can indeed be obtained from representatives of the form of eq.~(\ref{eq:twistF}).

Before moving on, however, some comments are in order regarding the product of twistor functions in eq.~(\ref{eq:CechDC}). As has been made clear repeatedly above, these are representatives of cohomology classes, and thus -- in a given theory -- can be subjected to equivalence transformations of the form of eq.~(\ref{fequiv}). However, the non-linear product that is needed to generate gravitational solutions in the twistor space double copy of eq.~(\ref{eq:CechDC}) (likewise in the four-dimensional case of refs.~\cite{White:2020sfn,Chacon:2021wbr}) is clearly incompatible with the ability to first perform equivalence transformations. This is not actually a problem if all one wants to do is to derive the Cotton double copy in position space: one merely regards the product as only being true for certain representatives in twistor space, such that any representatives which yield the correct double copy structure in position space (if it exists) will do. Nevertheless, it is desirable to have some motivation {\it a priori} for picking out certain representatives, where this would ideally relate to the physics of the double copy. Reference~\cite{Adamo:2021dfg} was the first to consider this point, using the language of Dolbeault rather than \u{C}ech cohomology. The authors considered certain radiative solutions, and showed that data at null infinity could be used to uniquely fix twistor representatives in the various theories entering the double copy. Reference~\cite{Chacon:2021lox} took a different approach, by looking at spacetime fields in Euclidean signature, and using existing the ideas of ref.~\cite{Woodhouse:1985id} to argue that upon choosing special cohomology representatives in twistor space (corresponding to harmonic differential forms), a product structure in twistor space is naturally obtained. Unfortunately, neither of these procedures is obviously related to the other, nor to the original BCJ double copy for scattering amplitudes. Reference~\cite{Guevara:2021yud}, however, provided a much better motivation for the formula of eq.~(\ref{eq:CechDC}), at least in principle, by showing that special twistor representatives can be defined by a certain integral transform acting on momentum-space amplitudes. Indeed, ref.~\cite{Luna:2022dxo} showed that these representatives are precisely those entering the twistor double copy of refs.~\cite{White:2020sfn,Chacon:2021wbr}.
Thus, the twistor double copy can indeed be viewed as arising from the BCJ double copy for three-point scattering amplitudes. Similar arguments can be used in the present context of solutions of topologically massive gauge theory and gravity, and we return to this in section~\ref{sec:cohomology}.

\subsection{Cotton double copy for Type N}

Let us now see how the type N Cotton double copy arises from twistor space. In line with our comments above, a type N solution should be generated by eq.~(\ref{eq:CechDC}), provided the gravity twistor representative has a simple pole in twistor space. We may thus choose representatives
\begin{equation}
    \check{f}_{-2}(Z^\alpha)=e^{-m\frac{\braket{a \mu}}{\braket{a \lambda}}}
    \frac{G_0(u,\lambda_A)}{\chi_{1}(u,\lambda_A)\xi_{1}(u,\lambda_A)},\quad
    \check{f}_{-4}^{(1,2)}(Z^\alpha)=e^{-m\frac{\braket{a \mu}}{\braket{a \lambda}}}
   \frac{G_0(u,\lambda_A)}{\chi_{1}(u,\lambda_A)(\xi_{1}(u,\lambda_A))^3},
    \label{typeNreps}
\end{equation}
where $\chi_{1}(u,\lambda_A)$ and $\xi_{1}(u,\lambda_A)$ are homogeneous of degree $1$ and have simple zeros and $G_0(u,\lambda_A)$ is homogeneous of degree $0$ and has no poles, such that one finds
\begin{equation}
\check{f}_{-6}(Z^\alpha)=e^{-m\frac{\braket{a \mu}}{\braket{a \lambda}}}
    \frac{G_0(u,\lambda_A)}{\chi_{1}(u,\lambda_A)(\xi_{1}(u,\lambda_A))^5}.
    \label{typeNreps2}
\end{equation}
Upon substituting this into the Penrose transform of eq.~(\ref{PenroseM}), we may carry out the latter by choosing the patch $U_0$ in eq.~(\ref{U0U1}), so that $\lambda_A=(1,z)$. We will assume without loss of generality that only the simple pole, which arises from $\xi_{1}$, of each cohomology representative lies in $U_0$. On general grounds, we may further define
\begin{align}
&\rho_x\left[\frac{\braket{a\mu}}{\braket{a\lambda}}\right]=q(x;z),\quad \rho_x\left[G_0(u,\lambda_A)\right]= G(x;z), \nonumber\\
    &\rho_x\left[\chi_{1}(u,\lambda_A)\right]=\frac{(z-z_0)}{N_1(x)},\quad
    \rho_x\left[\xi_{1}(u,\lambda_A)\right]=\frac{(z-z_1)}{N_2(x)} \ ,
    \label{AZdef}
\end{align}
where $z_0$ is the position of the simple zero in $\chi_{1}(u,\lambda_A)$, in terms of the parameter $z$, and the position dependence of each quantity arises upon imposing the incidence relation in eq.~\eqref{eq:incidence}. Equation~(\ref{PenroseM}) then becomes
\begin{align}
\phi_{AB\ldots D}(x)=&\frac{1}{2\pi i}\int_{\Gamma} \mathrm{d} z  (1,z)_{A} (1,z)_{B}\cdots (1,z)_{D}  G(x;z)  e^{-m\  q(x;z)}   \nonumber \\
&\quad\times \frac{N_1(x)}{z-z_0}\left(
\frac{N_2(x)}{z-z_1}
\right)^{2n+1}\nonumber\\
=&  (1,z_0)_{A} (1,z_0)_{B}\cdots (1,z_0)_{D}\, N_1(x) \,G(x;z_0)\,
\left(\frac{N_2(x)}{z_0-z_1}\right)^{2n+1}
e^{-m \ q(x;z_0)} \ ,
\label{phitypeN}
\end{align}
where we have carried out the contour integral in the second line by assuming that $q(x,z)$ is non-singular at $z=z_0$. The fields of eq.~(\ref{phitypeN}) clearly satisfy the Cotton double copy of Eq.~\eqref{eq:CottonDC}, when taking $C_{ABGD}=\phi_{ABGD}$ and $\Phi_{AB}=\phi_{AB}$. Thus, the Cotton double copy indeed emerges from a product in twistor space, as claimed.

To give an explicit example of the above construction, let us examine  pp-wave solutions, for which the following representatives in twistor space can be constructed for a spin-$n$ field:
\begin{equation}
  \check{f}_{-2-2n}=e^{-m\frac{\braket{a \mu}}{\braket{a \lambda}}} \frac{1}{\braket{o \lambda}}\left( \frac{1}{\braket{s \lambda}}\right)^{2n+1} g\left(\frac{u}{(s^A \lambda_A)^2}\right) \ .
\end{equation}
Here we have introduced the constant spinors $a^A=(1,0)$, $o^A=(0,1)$, and $s^A=(1,c)$, where $c\ \epsilon\ \mathbb{C}$. Comparing to eq.~\eqref{AZdef} we have
\begin{align}
&\rho_x\left[G_0(u,\lambda_A)\right]=\rho_x\left[g\left(\frac{u}{(s^A \lambda_A)^2}\right)\right]=g\left(\frac{(-t+y) z^2-2x z-y-t}{(1+c z)^2}\right) \nonumber \\
&\rho_x\left[\chi_{1}(u,\lambda_A)\right]=\rho_x\left[\braket{o \lambda}\right]=z,\quad
    \rho_x\left[\xi_{1}(u,\lambda_A)\right]=\rho_x\left[\braket{s \lambda}\right]=1+ c z \ .
\end{align}
where the pole $z_0=0$ is in $U_0$ and $z_1=-1/c$ is in $U_1$. Carrying out the relevant Penrose transforms as in eq.\eqref{phitypeN}, one finds
\begin{align}
C_{ABCD}=\phi(y_+,x)\ \alpha_{A} \alpha_{B}\alpha_{C} \alpha_{D} \ , \quad f_{AB}=\phi(y_+,x)\ \alpha_{A} \alpha_{B}\ , \quad  \phi(y_+,x)=g(y_+)e^{-m x} \ ,
\end{align}
with $y_\pm=t\pm y$ and $\alpha_A=(1,0)$ the principal spinor of the plane wave solutions. These are indeed the pp-wave solutions for topologically massive gravity, topologically massive electrodynamics, and a massive scalar field \cite{Chow:2009km}. 

\subsection{Beyond type N solutions}
\label{sec:beyondN}

Having reproduced the Cotton double copy for type N solutions of refs.~\cite{Gonzalez:2022otg,Emond:2022uaf}, it is natural to ask whether or not the arguments can be extended to type D solutions. The latter indeed double copy in the four-dimensional Weyl double copy, whose twistorial incarnation has been presented in refs.~\cite{White:2020sfn,Chacon:2021wbr}. The twistor description allows us to address this directly, and in fact shows that type D solutions do not obey a simple position space double copy in general. To see this, we may write the explicit formula
\begin{equation}
    \check{f}_{-2n-2}(Z^\alpha)=e^{-m\frac{\braket{a \mu}}{\braket{a \lambda}}} \frac{G_0(u,\lambda_A)}{(\chi_{1}(u,\lambda_A)\xi_{1}(u,\lambda_A))^{n+1}} \label{eq:twistF2} \ ,
\end{equation}
where as before, $\chi_{1}(u,\lambda_A)$ and $\xi_{1}(u,\lambda_A)$ are homogeneous of degree $1$ and have simple zeros, and $G_0(u,\lambda_A)$ is homogeneous of degree $0$. We can further define
\begin{equation}
    \rho_x\left[\frac{G_0(u,\lambda_A)}{(\chi_{1}(u,\lambda_A)\xi_{1}(u,\lambda_A))^{n+1}}\right]= \frac{N(x)}{(z-z_0)^{n+1}(z-z_1)^{n+1}},
    \label{Qform}
\end{equation}
where we have again imposed the incidence relation on the coordinate patch $U_0$ and we assume that only the pole at $z_0$ is in $U_0$. Upon substituting this into the Penrose transform, one finds
\begin{align}
\phi_{AB\ldots D}(x)=&\frac{1}{2\pi i}\int_{\Gamma} \mathrm{d} z  (1,z)_{A} (1,z)_{B}\cdots (1,z)_{D} \, G(x;z)
e^{-m\  q(x;z)} \nonumber\\
&\quad\times \left(\frac{N(x)}{(z-z_0)^{n+1}(z-z_1)^{n+1}}\right).
\label{massivePenrose}
\end{align}
For type D solutions, poles of second order or higher will be present in the integrand which, upon taking residues, will generate terms involving derivatives of the combination
\begin{displaymath}
    G(x;z)e^{-m q(x;z)}.
\end{displaymath}
While we have shown this explicitly for type D solutions, this behaviour will hold for all non-type N solutions. Thus, rather than a single term in position space, one will obtain a sum of terms, such that a simple product of spacetime fields is not obtained in general. One way of simplifying matters is to only consider twistor representatives such that the function $G(x;z)$ is constant. Indeed, all of the type D representatives considered in the four-dimensional twistor double copy of refs.~\cite{White:2020sfn,Chacon:2021wbr} were of this form. However, this will not suffice in the present context, due to the exponential factor $e^{-m q(x;z)}$, whose presence is an unavoidable consequence of considering topologically massive gauge and / or gravity theory. We therefore conclude that, unlike the case of the Weyl double copy in four spacetime dimensions, the exact position-space Cotton double copy will be restricted purely to type N solutions\footnote{Note that we have assumed that the factor in the exponential has no poles in $U_0$. Below we will see that for linearized solutions that can be constructed from three-point amplitudes, this is not the case. In such cases, we will again find that the position space Cotton double copy does not hold.}. 
This is in stark contrast to the case of the Weyl double copy in four spacetime dimensions, where some of the simplest relevant solutions -- consisting of simple point-like objects at the origin -- are of type D. We have thus explained why refs.~\cite{Gonzalez:2022otg,Emond:2022uaf} only succeeded in finding Cotton double copies for type N solutions. Our results are also interesting in that they show that, even for non-type N solutions, there can still be a simple product-like double copy structure in twistor space. The lack of a double copy in position space is a consequence of the generalised Penrose transform, and thus ultimately due to the presence of the topological mass.
It is instructive to illustrate the general discussion of this section with a concrete example. This is the subject of the following section.

\section{From scattering amplitudes to cohomology representatives}
\label{sec:cohomology}

In the previous section, we have seen that Cotton double copy follows naturally from minitwistor space, analogous to how the Weyl double copy in four spacetime dimensions can be derived using twistor methods~\cite{White:2020sfn,Chacon:2021wbr}. Until recently, quite how the Weyl (position-space) and twistor double copies related to the BCJ double copy for scattering amplitudes remained mysterious. This was first settled in refs.~\cite{Monteiro:2020plf,Monteiro:2021ztt}, which showed that the Weyl and BCJ double copy for scattering amplitudes are equivalent, where they overlap, by using the so-called {\it KMOC formalism}~\cite{Kosower:2018adc} that expresses classical solutions as inverse on-shell Fourier transforms of scattering amplitudes. Reference~\cite{Luna:2022dxo} investigated this further, by using methods developed in ref.~\cite{Guevara:2021yud} to show that one may split the inverse Fourier transform from momentum to position space into two stages. The first takes momentum-space scattering amplitudes into twistor space, thereby picking out a particular cohomology representative. The second comprises the Penrose transform from twistor to position space, and ref.~\cite{Luna:2022dxo} thus makes clear that the amplitude, twistor and Weyl double copies are precisely equivalent where they overlap. A canonical example is that of a point mass or charge in gravity / gauge theory respectively, corresponding to the well-known Schwarzschild and Coulomb solutions. Similar solutions exist in topologically massive theories, namely gravitational and gauge theory anyons, whose double copy properties have been explored in refs.~\cite{Burger:2021wss}. Such solutions are not type N, such that we do not expect them to possess a simple position-space double copy, according to the arguments of the previous section. However, we do expect to see a simple product-like twistor-space double copy, where the relevant cohomology representatives are picked out by scattering amplitudes in momentum space. It is interesting to confirm this by seeing what actually happens if we take the relevant scattering amplitudes, and generalise the arguments of refs.~\cite{Guevara:2021yud,Luna:2022dxo} to three-dimensional topologically massive gauge theories and gravity.

Let us begin by developing the necessary ideas from the KMOC formalism of ref.~\cite{Kosower:2018adc}, which must be adapted to the present context (see also refs.~\cite{Monteiro:2020plf,Monteiro:2021ztt,Emond:2022uaf} for relevant ingredients). We will first focus on a scalar field, which we can mode expand in the usual way as
\[
\phi(x) = \int d\Phi(q)\left[a(q)e^{-iq\cdot x} + a^\dagger(q)e^{iq\cdot x}\right],
\]
where 
\begin{equation}
\sd\Phi(q) = \frac{d^3q}{(2\pi)^3}\hat{\delta}(q^2+m^2)\Theta(q_0)
\label{dPhiq}
\end{equation}
is the three-dimensional on-shell measure and $\hat{\delta}(x)\equiv 2\pi \delta(x)$. We are interested in the field generated by a static particle of mass $M$, which we take to be described by an initial state
\[
\ket{\psi} = \int \sd\Phi(p)\psi(p)\ket{p},
\label{initialstate}
\]
where $\psi(p)$ is a wavefunction in momentum space, corresponding to a wavepacket sharply peaked around the classical momentum $p^\mu=M u^\mu$, with $u^\mu$ the 4-velocity. Evolving this state into the far future using the S-matrix, the classical field is given by the expectation value 
\[
\varphi(x) = \braket{\psi|S^\dagger\phi(x)S|\psi},
\label{phiexpect}
\]
which in turn yields
\[
\varphi(x) = \int \sd\Phi(q)\left[\braket{\psi|S^\dagger a(q)S|\psi}e^{-iq\cdot x} + h.c.\right]
\]
Next, we can adopt the conventional expansion of the $S$-matrix: 
\begin{equation}
S = 1+iT,
\label{Sexpand}
\end{equation}
and note that 
\begin{equation}
\braket{\psi|a|\psi}=0, 
\label{apsi0}
\end{equation}
given that there are no $\phi$ excitations in the initial state. We thus get
\[
\varphi(x) &= 2\Re i\int \sd\Phi(q)\sd\Phi(p)\sd\Phi(p')\psi(p)^*\psi(p')\left[\braket{p'|a(q)T|p}e^{-iq\cdot x}\right]\\
&= 2\Re i\int \sd\Phi(q)\sd\Phi(p)\hat{\delta}(2p\cdot q + q^2)\psi(p)^*\psi(p+q)\left[\cl{A}^{(3)}(q)e^{-iq\cdot x}\right],
\] 
where in the second line we have introduced the three-point amplitude for the emission of the $\phi$ field by the source:
\begin{equation}
\braket{p'|a(q)T|p} = \cl{A}^{(3)}(q)\hat{\delta}(p+p'-q).
\label{Ascal}
\end{equation}
To understand this equation, note that the $a(q)$ operator acts as a creation operator on the left, creating a quantum of the $\phi$ field. The expectation value of the T-matrix is then, by definition, the three-point amplitude multiplied by a momentum-conserving delta function. As shown in ref.~\cite{Kosower:2018adc}, by carefully accounting for factors of $\hbar$ (absent in natural units), one can neglect the shift by $q$ in the wavefunction, and also the $q^2$ term in the delta function. One may then integrate out the momentum $p$ by assuming that the wavefunction $|\psi(p)|^2$ is appropriately normalised to find
\[
\varphi(x) &= \frac{1}{M}\Re i\int \sd\Phi(q)\hat{\delta}(u\cdot q)\left[\cl{A}^{(3)}(q)e^{-iq\cdot x}\right].
\label{phiKMOC}
\]   
In words: the classical field is obtained as an on-shell inverse Fourier transform of the three-point amplitude. Following refs.~\cite{Guevara:2021yud,Luna:2022dxo}, we can split this transform into two stages as follows. First, we introduce spinor variables by appealing to eq.~(\ref{pAB}):
\[
q_{AB} = \omega(\lambda_{A}\bar{\lambda}_B + \lambda_{B}\bar{\lambda}_A).
\label{qAB}
\]
Here $\omega$ has units of energy, so that the spinors $(\lambda_A,\bar{\lambda}_B)$ are dimensionless, and defined only up to the little group scalings
\begin{equation}
    \lambda_A\rightarrow \xi \lambda_A,\quad \bar{\lambda}_B
    \rightarrow \frac{1}{\xi}\bar{\lambda}_B.
    \label{littlegroup}
\end{equation}
Transforming to the new variables, we find that
\[
d^3q = 2\omega^2d\omega|\braket{\lambda\bar{\lambda}}|\braket{\lambda d\lambda}\braket{\bar{\lambda} d\bar{\lambda}}.
\label{d3q}
\]
Furthermore, the three-point amplitude for a scalar field emitted by a scalar source is simply given by a coupling constant, which we set to unity in what follows. Equation~(\ref{phiKMOC}) then becomes
\[
\varphi(x) = \Re \frac{i}{2\pi M}\int d\omega \braket{\lambda d\lambda}\braket{\bar{\lambda}d\bar{\lambda}}|\omega|\, |\langle \lambda\bar{\lambda}\rangle|\, \delta(\braket{\lambda|u|\bar{\lambda}})\,\delta\left(\omega^2\braket{\lambda\bar{\lambda}}^2 + m^2\right)e^{-i\omega\braket{\lambda|x|\bar{\lambda}}},
\label{phiKMOC2}
\]
where we have chosen to work in the rest frame where $u^\mu=(1,0,0)$ and thus $q_0=0$, and we have used the fact that $\Theta(q_0)=\Theta(0)=1/2$. The first delta function then implies that \[
\ket{\bar{\lambda}} \propto u\ket{\lambda},
\]
which we may write as an equality by relying upon the little group rescaling of eq.~(\ref{littlegroup}) (see also ref.~\cite{Luna:2022dxo} for a discussion of this point). Performing the $\bar{\lambda}$ integral then yields
\begin{equation}
    \phi=\Re \frac{i}{2\pi M}\int d\omega \braket{\lambda d\lambda}
    |\omega|\,|\langle \lambda|u|\lambda\rangle|\,\delta(
    \omega^2\langle\lambda|u|\lambda\rangle^2+m^2)
    e^{-i\omega\langle\lambda|xu|\lambda\rangle}.
    \label{phitransform1}
\end{equation}
To make further progress, we note that in our original metric signature $(-,+,+)$ for $m \in \R$, it is impossible to simultaneously solve the dual kinematic conditions
\begin{equation}
    u\cdot q = 0,\quad q^2+m^2=0.
    \label{kinconds}
\end{equation}
 To get around this problem, we may instead analytically continue to $(+,+,-)$ signature, by setting
\begin{equation}
    (q_0,q_2)=i(\tilde{q}_0,\tilde{q}_2),\quad (t,y)=i(\tilde{t},
    \tilde{y}).
    \label{qcontinue}
\end{equation}
From eq.~(\ref{pAB}) applied to $q^\mu$ with $\tilde{q}_0=0$, we may rescale according to eq.~(\ref{littlegroup}) to write
\begin{equation}
    \lambda_A=(1,z),\quad z=\frac{q_1-im}{i\tilde{q}_2}.
    \label{lambdaparam}
\end{equation}
The analytically continued kinematic constraint
\begin{equation}
    \tilde{q}_1^2-\tilde{q}_2^2+m^2=0
    \label{kincond2}
\end{equation}
then implies $|z|=1$, and we also have
\begin{displaymath}
\langle \lambda |u|\lambda\rangle\rightarrow i\langle \lambda |u|\lambda\rangle,\quad
\langle \lambda |xu|\lambda\rangle\rightarrow i\langle \lambda |xu|\lambda\rangle,
\end{displaymath}
so that eq.~(\ref{phitransform1}) becomes
\begin{equation}
       \phi=\Re \frac{i}{4\pi M m}\int d\omega \braket{\lambda d\lambda}
    |\omega|\left[\delta\left(\omega-\frac{m}{\langle\lambda |u|\lambda\rangle}\right)+\delta\left(\omega+\frac{m}{\langle\lambda |u|\lambda\rangle}\right)\right]
    e^{\omega\langle\lambda|xu|\lambda\rangle)},
    \label{phitransform2} 
\end{equation}
where we have used the standard identity
\begin{equation}
    \delta(x^2-\alpha^2)=\frac{1}{2|\alpha|}\Big[\delta(x-\alpha)
    +\delta(x+\alpha)\Big].
    \label{deltaid}
\end{equation}
It turns out that both delta function contributions are the same, such that we may take only the first with a factor of two. We then arrive at
\begin{equation}
\varphi(x)=\Re \frac{i}{M}\int_{\rm \Gamma}\frac{\langle \lambda
d\lambda\rangle}{2\pi}\frac{e^{m\frac{\langle\lambda|x u|
\lambda\rangle}{\langle\lambda|u|\lambda\rangle}}}
{\langle \lambda|u|\lambda\rangle},
\label{PenroseKMOC}
\end{equation}
where $\Gamma$ is the appropriate integration contour. Recognising $\langle \lambda |x=\langle\mu|$, eq.~(\ref{PenroseKMOC}) has precisely the form of the Penrose transform integrand of eq.~(\ref{eq:solution f}), where in this case $|a\rangle=u|\lambda\rangle$, and a specific function of $\lambda_A$ occurs, due to having transformed a particular momentum-space amplitude into (mini-)twistor space. This is a highly useful validation that eq.~(\ref{eq:solution f}) is the correct Penrose transform integrand for topologically massive theories. But it also forms the basis for calculating similar results for gauge and gravity anyon solutions, and examining their double copy properties. 

Explicit calculation of the various spinor products in eq.~(\ref{PenroseKMOC}) yields 
\[
\phi(x) = \Re \frac{i}{M} \int_{\Gamma}\frac{d z}{2\pi}\frac{e^{-m\frac{-2i\tilde{y}z + x(1-z^2)}{1+z^2}}}{1+z^2},
\label{phiint}
\]
such that transforming to the variable $z=e^{i\alpha}$ gives
\begin{equation}
\phi(x)=-\frac{1}{4\pi M}\Re \int_{-\pi/2}^{\pi/2}\frac{d\alpha}
{\cos\alpha}\exp\left(im\frac{\tilde{y}-x\sin\alpha}{\cos\alpha}
\right),
    \label{phitransform4}
\end{equation}
We can use the rotational symmetry of the scalar solution to set $\tilde{y}=0$ and $x=r$, such that eq.~(\ref{phitransform4}) becomes
\begin{align}
\phi=-\frac{1}{4\pi M}\Re\int_{-\pi/2}^{\pi/2}d\alpha 
\frac{e^{-imr\tan\alpha}}{\cos\alpha}=
-\frac{1}{2\pi M}K_0(mr),
    \label{phires}
\end{align}
where we have used a known integral representation for the modified Bessel function of the second kind~\footnote{Strictly speaking, the argument of the $K_0$ function should be $|mr|$, but $m$ and $r$ are both positive in our case. This justifies the remark made above that the contributions from both delta functions in eq.~(\ref{phitransform2}) give the same result.}. Finally we may analytically continue back to the metric signature $(-,+,+)$, which does not change the form of the result, but ensures that $r=\sqrt{x^2+y^2}$.

\subsection{Topologically massive gauge theory}

In the previous section, we have seen that a classical scalar field can be obtained as an inverse on-shell Fourier transform of a three-point amplitude, such that splitting this transform into two stages allows us to recognise the Penrose transform from twistor to position space. Similar arguments apply to topologically massive gauge theory and gravity, and thus allow us to examine how to Cotton double copy does or does not work for pointlike solutions. In the gauge theory, we are concerned with the relevant curvature spinor, derived from the dual field strength in the case of topologically massive electromagnetism. This is in turn related to the three-point amplitude for emission of a photon by a source~\cite{Emond:2022uaf}, which we again take to be a scalar particle of mass $M$.
 
We can mode expand the dual field strength on-shell in the usual way\footnote{We note however that there is no sum over helicities here, since topologically massive theories propagate only a single helicity.}
\[
\tilde{F}^\mu(x) = \frac12\epsilon^{\mu\nu\rho}F_{\nu\rho}= -im\int \sd\Phi(q)\left[a(q)\epsilon^\mu(q)e^{-iq\cdot x} + a^\dagger(q)\epsilon^{*\mu}(q)e^{iq\cdot x}\right].
\]
Then, analogously to the scalar case of eq.~(\ref{phiexpect}), the curvature spinor is related to the expectation value of the field strength evolved into the far future by the $S$-matrix 
\[
\varphi_{AB}(x) = \braket{\psi|S^\dagger\tilde{F}^\mu(x)\sigma_{\mu AB}S|\psi},
\]
where $\ket{\psi}$ is the initial one-particle state defined in eq.~(\ref{initialstate}), and where we now assume this particle is charged such that it may emit photons. The electromagnetic curvature spinor is then given by
\[
\varphi_{AB}(x) = -im\int \sd\Phi(q)\left[\braket{\psi|S^\dagger a(q)S|\psi}\epsilon_{AB}(q)e^{-iq\cdot x} + h.c.\right],
\]
such that upon following similar steps to those leading to eq.~(\ref{phiKMOC}), we arrive at
\[
\varphi_{AB}(x) &= -\frac{m}{M}\Re\int \sd\Phi(q)\hat{\delta}(u\cdot q)\left[\cl{A}_{\rm gauge}^{(3)}(q)\epsilon_{AB}(q)e^{-iq\cdot x}\right],
\label{phiABcalc}
\]   
where ${\cal A}_{\rm gauge}$ is the appropriate three-point amplitude for the emission of a (topologically massive) gauge boson from a scalar. As in the scalar case, we may transform to spinor coordinates according to eq.~(\ref{qAB}). A suitable choice for the polarization vector is 
\[
\epsilon^\mu = \omega\frac{\braket{\lambda|\gamma^\mu|\lambda}}{2m}~~~\implies~~~\epsilon_{AB} = \frac{\omega}{m}\lambda_A\lambda_B,
\label{epsilondef}
\]
Furthermore, the amplitude ${\cal A}^{(3)}_{\rm gauge}$ can be fixed by dimensional analysis and little group scaling~\cite{Moynihan:2020ejh}:
\[
\cl{A}^{(3)} = 2eM(u\cdot\epsilon(q)). \label{eq:A3gauge}
\]
Repeating similar arguments to those leading to eq.~(\ref{PenroseKMOC}), we ultimately find
\[
\varphi_{AB}(x) = -em\Re i\int_\Gamma \frac{\braket{\lambda \sd\lambda}}{2\pi}\lambda_A\lambda_B \frac{e^{m\frac{\braket{\mu|u|\lambda}}{\braket{\lambda|u|\lambda}}}}{\braket{\lambda|u|\lambda}^2}.
\label{phiABKMOC}
\]
This looks almost identical to the scalar case of eq.~(\ref{phiKMOC}), but such that the integrand now contains additional powers of spinor variables, as is appropriate for a spin-1 field. Again, we have obtained a specific example of the massive Penrose transform integrand of eq.~(\ref{eq:solution f}). 

Equation~(\ref{phiABKMOC}) will be useful for examining double copy properties of anyon solutions, but let us first carry out the Penrose transform to position space. To do this, we may choose the parametrisation $\lambda_A=(1,z)$, and define the master integral
\begin{equation}
    I_{p,q}(x,y)=\Re i\int_\Gamma \frac{dz}{2\pi} \frac{z^p\,e^{\frac{- m[-2yz+x(1-z^2)]}{1+z^2}}}
    {(1+z^2)^{q+1}}.
    \label{Ipqdef}
\end{equation}
In terms of this integral, the scalar field of eq.~(\ref{phiint}) can be written as 
\begin{equation}
    \phi=\frac{1}{M} I_{0,0},
    \label{phi00}
\end{equation}
after analytically continuing back to the mostly plus metric signature, from which we find
\begin{equation}
    I_{0,0}=A K_0(m r),\quad A=-\frac{1}{\pi}.
    \label{I00res}
\end{equation}
Similarly, the field strength spinor of eq.~(\ref{phiABKMOC}) has the form
\begin{equation}
    \varphi_{00}(x)=-em I_{0,1},\quad 
    \varphi_{01}=\varphi_{10}=-emI_{1,1},\quad
    \varphi_{11}=-emI_{2,1}.
    \label{varphiI}
\end{equation}
Given the result of eq.~(\ref{I00res}), we may carry out the integrals on the right-hand side of eq.~(\ref{varphiI}) without performing any further explicit integrals. To see how, note that we can differentiate eq.~(\ref{Ipqdef}) to obtain the recurrence relations
\begin{equation}
    \partial_x I_{p,q}=- m[I_{p,q+1}-I_{p+2,q+1}],\quad
    \partial_y I_{p,q}=2 m I_{p+1,q+1}.
    \label{recurrence1}
\end{equation}
A further recurrence relation can be obtained by substituting 
\begin{displaymath}
1=\frac{1+z^2}{1+z^2}
\end{displaymath}
into the integrand of eq.~(\ref{Ipqdef}), yielding
\begin{equation}
    I_{p+2,q+1}=I_{p,q}-I_{p,q+1}.
    \label{recurrence2}
\end{equation}
We thus find that all integrals entering the field strength spinor of eq.~(\ref{varphiI}) can be expressed in terms of derivatives of eq.~(\ref{I00res}), leading to the explicit results
\begin{align}
    I_{0,1}&=\frac{A}{2}\left[K_0(mr)
    +\frac{x}{ r}K_1( m r)\right];\notag\\
    I_{1,1}&=-\frac{Ay}{2r}K_1(m r);\notag\\
    I_{2,1}&=\frac{A}{2}\left[K_0(mr)
    -\frac{x}{r}K_1(m r)\right].
    \label{I1results}
\end{align}
We thus find 
\begin{align}
    \varphi_{00}&=\frac{em}{2\pi}
    \left[K_0(mr)
    +\frac{x}{r}K_1(m r)\right];\notag\\
    \varphi_{01}&=-\frac{em}{2\pi}\frac{y\,K_1(mr)}{r}\notag\\
    \varphi_{11}&=\frac{em}{2\pi}
    \left[K_0(mr)
    -\frac{x}{r}K_1(m r)\right].
    \label{varphisol}
\end{align}

\subsection{Topologically massive gravity}

A similar analysis to the previous section may be carried out for topologically massive gravity, where we may consider a scalar particle emitting (massive) gravitons. The classical result for the Cotton spinor of eq.~(\ref{Cottonspinor}) is then given by 
\begin{equation}
    C_{ABCD}(x)=\langle\psi|S^\dag C_{ABCD} S|\psi\rangle,
    \label{Cexpect}
\end{equation}
such that the analogue of eq.~(\ref{phiABcalc}) is, following \cite{Emond:2022uaf}
\begin{flalign}
    C_{ABCD}(x)=-\kappa\frac{m}{2M}\Re i\int d\omega &|\braket{\lambda\bar{\lambda}}|\frac{\braket{\lambda d\lambda}\braket{\bar{\lambda} d\bar{\lambda}}}{2\pi}\delta(\braket{\lambda|u|\bar{\lambda}})\delta(\omega^2 \braket{\lambda\bar{\lambda}}^2 + m^2)
    |\omega|^3\nn\\ &\times \left[{\cal A}_{\rm grav.}(q)\lambda_{A}\lambda_B \lambda_C \lambda_De^{-iq\cdot x}
    \right].
    \label{Ccalc}
\end{flalign}

In eq.~(\ref{Ccalc}), ${\cal A}_{\rm grav.}$ is the three-point amplitude for emission of a graviton by a scalar. To find this, we may quote a general result for a spin-$s$ field coupled to two scalars:
\[
\cl{A}^{(3)}_{+s} = g_s M^s (2u\cdot\epsilon(q))^s,
\]
where $g_s$ is some coupling. By similar arguments to the gauge theory case, this simply evaluates to a constant, such that the analogue of eq.~(\ref{phiABKMOC}) turns out to be
\begin{equation}
    C_{ABCD}(x)=-\frac{\kappa^2m^3 M}{2}\Re i\int_\Gamma \frac{\langle \lambda d\lambda\rangle}
    {2\pi} \lambda_A \lambda_B \lambda_C \lambda_D 
    \frac{e^{m\frac{\mu|u|\lambda\rangle}{\langle \lambda|u|\lambda\rangle}}}{\langle \lambda|u|\lambda\rangle^3}.
    \label{CABCDint}
\end{equation}

In terms of the basis of integrals defined in eq.~(\ref{Ipqdef}), we then have 
\begin{equation}
C_{ABCD}=-\frac{\kappa^2m^3 M}{2}I_{n_1(ABCD),2},
    \label{Ccalc1}
\end{equation}
where $n_1(ABCD)$ is the number of $1$ indices (rather than $0$ indices) in the string $ABCD$. By use of the above recurrence relations we find
\begin{align}
    C_{0000}&=\frac{\kappa^2m^3M}{8\pi}\left[\left(2-\frac{y^2}{r^2}\right) K_0(mr)
    +\left(\frac{2x}{r}+\frac{(r^2-2y^2)}{mr^3}\right)K_1(mr)\right];\notag\\    
    C_{1000}&=\frac{\kappa^2m^3M}{8\pi}\left[
    -\frac{xy}{r^2}K_0(mr)+\left(-\frac{y}{r}-\frac{2xy}{mr^3}
    \right)K_1(mr)\right];\notag\\
    C_{1100}&=\frac{\kappa^2m^3M}{8\pi}\left[\frac{y^2}{r^2}K_0(mr)+\left(
    -\frac{1}{mr}+\frac{2y^2}{mr^3}\right)K_1(mr)\right];\notag\\
    C_{1110}&=\frac{\kappa^2m^3M}{8\pi}\left[
    \frac{xy}{r^2}K_0(mr)+\left(
    -\frac{y}{r}+\frac{2xy}{mr^3}
    \right)K_1(mr)\right];\notag\\
    C_{1111}&=\frac{\kappa^2m^3M}{8\pi}\left[
    \frac{(2r^2-y^2)}{r^2}K_0(mr)
    +\left(-\frac{2x}{r}+\frac{1}{mr}-\frac{2y^2}{mr^3}
    \right)K_1(mr)\right].
    \label{CABCDresults}
\end{align}
We have checked explicitly that the results of eqs.~(\ref{varphisol}, \ref{CABCDresults}) agree with the known anyon solutions in topologically massive gauge and gravity theory~\cite{Deser:1989ri}, once these are translated into the spinorial language, and up to an overall normalisation constant (which we have defined differently in our choice of constant amplitudes above). 

\subsection{Double copy properties of anyon solutions}
\label{sec:anyonDC}

In the previous sections, we have seen that the solution for spin-$n$ field for a pointlike source, in a (topologically) massive theory, takes the general form
\begin{equation}
    \psi_{AB\ldots D}(x)=K_n\int_\Gamma \frac{\langle \lambda d\lambda\rangle}
    {2\pi} \lambda_A \lambda_B \ldots \lambda_D 
    \frac{e^{m\frac{\mu|u|\lambda\rangle}{\langle \lambda|u|\lambda\rangle}}}{\langle \lambda|u|\lambda\rangle^{n+1}},
    \label{psiABD}
\end{equation}
where $K_n$ is a normalisation constant. Each solution is a special case of the massive Penrose transform of eq.~(\ref{massivePenrose}), and we thus see that there is a multiplicative double-copy structure in twistor space. That is, upon picking out a cohomology representative $f_n(\lambda_A, u)$ for each spin-$n$ field (scalar, gauge and gravity) by transforming amplitudes into twistor space, these representatives are related by a simple multiplicative rule
\begin{equation}
    f_2(\lambda_A,u)=\frac{f_1(\lambda_A,u)f_1(\lambda_A,u)}{f_0(\lambda_A,u)}.
    \label{twistorDC}
\end{equation}
The same result is obtained in four spacetime dimensions~\cite{Luna:2022dxo}. However, unlike in that case, the simple multiplicative nature of the double copy in twistor space does not correspond to a simple structure in position space, as comparison of eqs.~(\ref{varphisol}, \ref{CABCDresults}) makes clear. This thus provides an explicit illustration of the general discussion in section~\ref{sec:beyondN}, namely that the presence of the exponential factor in the massive Penrose transform disrupts the simple nature of the position-space double copy. Given that fields generated by a pointlike source are perhaps the simplest static solutions one can imagine, this bolsters the conclusions of ref.~\cite{Luna:2022dxo}, that exact position space double copies are rather special and restricted in nature, and that the double copy prefers to live in momentum space. Another interesting feature of the three-dimensional solutions considered here is that, in contrast to their four-dimensional counterparts, the twistor space representatives have essential singularities, rather than poles. Again this is due to the presence of the exponential factor, and guarantees that one is able to reconstruct the relevant transcendental functions entering the spacetime field (i.e. modified Bessel functions) upon taking residues in the Penrose transform integral.

\section{Conclusion}
\label{sec:discuss}

In this paper, we have examined the Cotton double copy recently presented in refs.~\cite{Gonzalez:2022otg,Emond:2022uaf}, that relates solutions of topologically massive gauge and gravity theories. It is a three dimensional (massive) counterpart of the Weyl double copy for certain exact solutions in four spacetime dimensions~\cite{Luna:2018dpt}. However, whereas the latter is known to apply to arbitrary Petrov type D vacuum solutions, the former is apparently restricted solely to the type N case. In order to clarify this issue, we have here used twistor methods, which have previously been useful in examining the origin (and special nature) of the Weyl double copy~\cite{White:2020sfn,Chacon:2021wbr,Chacon:2021hfe,Luna:2022dxo}. 

In three spacetime dimensions, the relevant twistor space is called {\it minitwistor space}, and we have reviewed known results from the twistor literature~\cite{tsai_1996} that provide a massive generalisation of the well-known Penrose transform relating classical fields in spacetime with cohomology classes in twistor space. Armed with this Penrose transform, one may show that although it is possible to construct gravitational cohomology representatives by combining representatives from (massive) scalar and gauge theories, this leads to a simple position-space double copy only in the case of type N solutions. We thus confirm the results of refs.~\cite{Gonzalez:2022otg,Emond:2022uaf}, and further clarify our results by considering arguably the simplest possible static solutions, corresponding to a pointlike source. The relevant classical fields can be expressed as on-shell inverse Fourier transforms of three-point amplitudes, following the methods of ref.~\cite{Kosower:2018adc}. By splitting this transform into two stages, we first transform amplitudes into minitwistor space, revealing that the remaining to spacetime takes precisely the form of the massive Penrose transform mentioned above. Although we find a simple multiplicative double copy in twistor space, this fails to translate to a simple relationship in spacetime, thus validating our more general analysis. 

The emerging picture from this and similar recent studies~\cite{Luna:2022dxo} is that exact position-space double copies are rare. However, knowing that they exist -- and what their limitations are -- is undoubtedly useful. Methods for elucidating the landscape of exact double copies are a necessary part of this ongoing effort, and we hope that the twistor methods developed in this paper may prove of further use in this regard.

\section*{Acknowledgments}

We are grateful to Tim Adamo and Graham Brown for useful discussions. This work has been supported by the UK Science and Technology Facilities Council (STFC) Consolidated Grant ST/P000754/1 ``String theory, gauge theory and duality". \\
MCG is supported by
the European Union’s Horizon 2020 Research Council grant 724659 MassiveCosmo ERC–2016–COG and the STFC grants ST/P000762/1 and ST/T000791/1. JR is supported by the National Science and Technology Council of Taiwan grant NSTC 111-2811-M-002-125. NM is supported by STFC grant ST/P0000630/1 and the Royal Society of Edinburgh Saltire Early Career Fellowship. WTE is supported by the Czech Science Foundation GACR, project 20-16531Y.
\bibliographystyle{JHEP}
\bibliography{topgrav}	
\end{document}